\documentclass[twocolumn,secnumarabic,amssymb,superscriptaddress, showpacs, aps, prd]{revtex4-2}

\usepackage[english]{babel}
\usepackage{graphicx}
\usepackage{bm}
\usepackage{amsmath}
\allowdisplaybreaks
\usepackage{dsfont} 
\usepackage{mathtools} 
\usepackage{amsfonts}
\usepackage{xargs} 
\usepackage{ifmtarg} 
\usepackage{physics} 
\usepackage{siunitx}
\usepackage[acronym]{glossaries}

\newcommand{\derd}{\operatorname{d}}
\newcommand{\myindex}[1]{{\scriptscriptstyle #1}}
\newcommand{\exposant}[1]{{\scriptscriptstyle #1}}
\newcommand{\myexpr}[3]{\bm{#1}_\myindex{#2}^\exposant{#3}}
\newcommand{\myexprlower}[2]{#1_\myindex{#2}}

\newcommandx{\trafo}[4][2=,4=]{\myexpr{T}{\mathcal{#3}_\myindex{#4}}{\mathcal{#1}_\myindex{#2}}}
\newcommand{\mymat}[1]{\bm{#1}}
\newcommand{\trafomat}[1]{\mymat{T}_{#1}}
\newcommandx{\myvecwithind}[2][2=]{\bm{#1}_\myindex{#2}}
\newcommand{\myvec}[1]{\bm{#1}}
\newcommandx{\scalwithref}[7][3=,4=,6=,7=]{#1_\myindex{\mathcal{#2}_\myindex{#3}^\exposant{#4}/\mathcal{#5}_\myindex{#6}^\exposant{#7}}}
\newcommandx{\vecwithrefrel}[7][3=,5=,7=]{\myexpr{#1}{\mathcal{#2}_\myindex{#3} / \mathcal{#4}_\myindex{#5} }{\mathcal{#6}_\myindex{#7}}}
\newcommandx{\vecwithref}[5][3=,5=]{\myexpr{#1}{\mathcal{#2}_\myindex{#3}}{\mathcal{#4}_\myindex{#5}}}
\newcommandx{\momentinertia}[5][3=,5=]{\myexpr{I}{\text{#1}/#2_\myindex{#3}}{\mathcal{#4}_\myindex{#5}}}

\newcommand{\sint}[1]{\sin{#1}}
\newcommand{\cost}[1]{\cos{#1}}

\newcommandx{\framevu}[3][3=]{\vu{e}_{#2,\mathcal{#1}_\myindex{#3}}}

\newacronym{e2e}{E2E}{End-To-End}
\newacronym{inrep}{INREP}{Initial Noise REduction Pipeline}
\newacronym{tdi}{TDI}{Time-Delay Interferometry}
\newacronym{ttl}{TTL}{Tilt-To-Length}
\newacronym{dfacs}{DFACS}{Drag-Free and Attitude Control System}
\newacronym{ldc}{LDC}{LISA Data Challenge}
\newacronym{ldpg}{LDPG}{LISA Data Processing Group}
\newacronym{lsg}{LSG}{LISA Science Group}
\newacronym{lig}{LIG}{LISA Instrument Group}
\newacronym{simwg}{LDPG-SimWG}{Simulation Working Group}
\newacronym{lisa}{LISA}{Laser Interferometer Space Antenna}
\newacronym{siso}{SISO}{Single-Input Single-Ouput}
\newacronym{mimo}{MIMO}{Multiple-Input Multiple-Output}
\newacronym{emri}{EMRI}{Extreme Mass Ratio Inspiral}
\newacronym{ifo}{IFO}{Interferometry System}
\newacronym{grs}{GRS}{Gravitational Reference System}
\newacronym{dws}{DWS}{Differential Wavefront Sensing}
\newacronym{mosa}{MOSA}{Moving Optical Sub-Assembly}
\newacronym{mps}{MPS}{Micro Propulsion System}
\newacronym{eom}{EoM}{Equations of Motion}
\newacronym{com}{CoM}{center of mass}
\newacronym{gw}{GW}{gravitational wave}
\newacronym{isi}{ISI}{inter-spacecraft interferometer}
\newacronym{tmi}{TMI}{test-mass interferometer}
\newacronym{ldws}{LDWS}{Long-arm Differential Wavefront Sensing}
\newacronym[\glslongpluralkey={Degrees of Freedom}]{dof}{DoF}{Degree of Freedom}
\newacronym{icrs}{ICRS}{International Celestial Reference System}
\newacronym{rk4}{RK4}{Runge-Kutta 4}
\newacronym{rkf45}{RKF45}{Runge-Kutta-Fehlberg 4 (5)}
\newacronym{lte}{LTE}{Local Truncation Error}
\newacronym{gte}{GTE}{Global Truncation Error}
\newacronym[\glslongpluralkey={Test Masses}]{tm}{TM}{Test Mass}
\newacronym{lti}{LTI}{Linear Time-Invariant}
\newacronym{asd}{ASD}{Amplitude Spectral Density}
\newacronym{psd}{PSD}{Power Spectral Density}
\newacronym{rms}{RMS}{Root Mean Squared}
\newacronym{ls}{LS}{Least Squares}
\newacronym{ols}{OLS}{Ordinary Least Squares}
\newacronym{esa}{ESA}{European Space Agency}
\newacronym[\glsshortpluralkey={s/c}]{sc}{s/c}{spacecraft}
\newacronym{adc}{ADC}{Analogue-to-Digital Converter}
\newacronym{df}{DF}{Drag-Free}
\newacronym{bam}{BAM}{Beam-Alignment Mechanism}
\newacronym{mcmc}{MCMC}{Markov chain Monte Carlo}

\usepackage[pdfborderstyle={/S/U/W 1}]{hyperref}
\newcommand{\myhyperref}[1]{\hyperref[#1]{\ref{#1}}}

\begin{document}

\title{LISA Non-Linear Dynamics and Tilt-To-Length Coupling}%

\author{Lavinia Heisenberg}
\affiliation{Institut f\"ur Theoretische Physik, Universit\"at Heidelberg, Philosophenweg 16, 69120 Heidelberg, Germany}

\author{Henri Inchausp\'e}
\affiliation{Institute for Theoretical Physics, KU Leuven, Celestijnenlaan 200D, B-3001 Leuven, Belgium}
\affiliation{Leuven Gravity Institute, KU Leuven, Celestijnenlaan 200D box 2415, 3001 Leuven, Belgium}

\author{Sarah Paczkowski}
\affiliation{Max Planck Institute for Gravitational Physics (Albert Einstein Institute), Callinstr. 38, 30167 Hannover, Germany}
\affiliation{Leibniz Universit\"at Hannover, Callinstr. 38, 30167 Hannover, Germany}

\author{Ricardo Waibel}
\email[Corresponding author: ]{waibel@thphys.uni-heidelberg.de}
\affiliation{Institut f\"ur Theoretische Physik, Universit\"at Heidelberg, Philosophenweg 16, 69120 Heidelberg, Germany}

\date{January 2026}
\begin{abstract}
    For the LISA mission, Tilt-To-Length (TTL) coupling is expected to be one of the dominant instrumental noise contributions after laser frequency noise is suppressed based, on assumptions on the size of the coupling and angular jitter levels. This work uses for the first time a closed-loop, non-linear, and time-varying dynamics implementation to simulate detailed angular jitters for the spacecraft and optical benches in LISA. In turn, this gives an improved expectation of the TTL contribution to the interferometric output. It is shown that the TTL coupling impact is limited given current estimates on the size of coupling coefficients.
    A time-domain Least Squares estimator is used to infer the TTL parameters from the simulated measurements. The bias and correlations limit the estimator in the case of regular datasets with amplified TTL coefficients to a relative error of $10\%$, but the subtraction of the TTL signal still works well. For lower readout noises, the estimation error diverges, which can be mitigated using a regularization term. Alternatively, using sinusoidal maneuvers improves the inference to a high accuracy of $0.1\%$ for TTL coefficients around the expected level, removing all correlations in the inferred parameters. This validates the maneuver design by Wegener et al. \cite{wegener2025design} in this closed-loop setting.
\end{abstract}

\maketitle

\section{Introduction}
The \gls{lisa} is a mission led by \gls{esa} to detect \glspl{gw} in the millihertz regime using a space-based detector \cite{colpi2024lisadefinitionstudyreport}. \gls{lisa} was adopted in 2024, as the third large-class mission, L3, and is currently planned to launch in 2035. This newly available frequency band from \SI{0.1}{\milli\Hz}--\SI{1}{\Hz} of \glspl{gw} has interesting astrophysical and cosmological sources. For example massive black hole binaries are expected to be detected with high signal-to-noise ratios. Other sources include galactic binary systems, extra-galactic stellar-mass binaries, and possibly stochastic signals from the early universe, such as cosmic strings or signals from first-order phase transitions. These sources are then expected to give answers to important questions on the nature of gravity, the formation history of black holes, the expansion of our universe, and possibly give insights into the early stages of the universe \cite{colpi2024lisadefinitionstudyreport}. This mission will complement already operating observatories on the ground \cite{abbott2020prospects,aasi2015advanced,acernese2014advanced,akutsu2021overview,dooley2016geo}, operating in the \SI{10}{\Hz}--\SI{10}{\kilo\Hz} regime, and using pulsar-timing-arrays \cite{agazie2023nanograv,antoniadis2023second,reardon2023search,xu2023searching}, accessing the frequencies with \SI{1}{\nano\Hz}--\SI{10}{\micro\Hz}.

\begin{figure}[!ht]
\includegraphics[width=.95\linewidth]{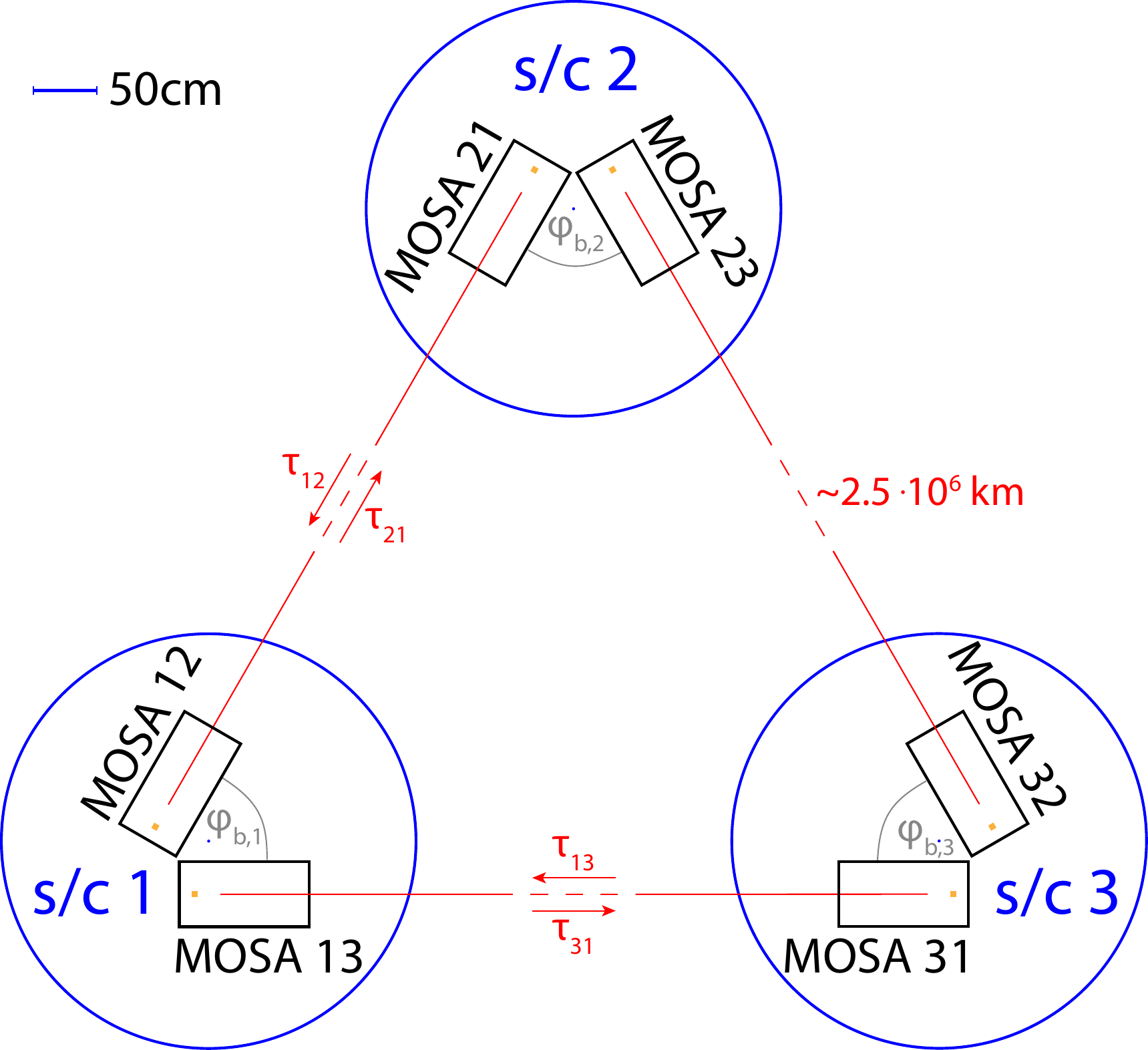}
\caption{Figure cut of constellation geometry of \gls{lisa} in the triangle plane. The scheme for numbering \gls{sc}, \glspl{mosa}, light travel times $\tau$, and \gls{mosa} opening angles $\varphi_b$ are defined.}
\label{fig:constellation}
\end{figure}

The design uses three \gls{sc} in an approximately equilateral formation, separated by about \num{2.5} million kilometers \cite{colpi2024lisadefinitionstudyreport}. \Glspl{tm} on board of \gls{sc} provide an inertial reference system, which is then measured using a laser system. Every \gls{sc} contains two \glspl{mosa}, each containing a laser system capable of sending a beam to another \gls{sc} and able to receive one. These established links are then one crucial part of the interferometric measurement, the second part being the local interferometric measurements of the \gls{tm} position within the \gls{sc}. Figure~\myhyperref{fig:constellation} illustrates this constellation geometry and introduces the numbering conventions for the $3$ \gls{sc} and $6$ \glspl{mosa}. Similar detector concepts have also been proposed \cite{luo2016tianqin,luo2021taiji}, most results apply to the whole class. Here we will, however, focus on the concrete design of \gls{lisa}.

Operating such a detector in space presents unique challenges, such as achieving picometer-level measurements across millions of kilometers and femto-Newton-level free fall of the \glspl{tm} in the frequency band. While the LISA Pathfinder mission already demonstrated the required performance for the free-fall and the local interferometry \cite{armano2016sub,armano2018beyond,lpf_sensor_noise21}, inter-satellite interferometry separated by approximately \SI{220}{\kilo\m} was demonstrated by the GRACE Follow-on mission \cite{grace_fo_ifo_19}. Additionally, the typical Michelson-interferometer approach known from the ground-based detector cannot be used directly. This is because the received laser beam power is too low to be reflected directly and because the constellation triangle will drift over time, which means that the dominant laser phase/frequency noise does not automatically cancel as is the case in an equal-arm-length interferometer. For this, post-processing techniques of the beatnote measurements of the interferometers known as \gls{tdi} have been developed that can successfully suppress this dominant noise \cite{tinto_time-delay_2020,muratore_revisitation_2020,bayle2021adapting}.

In the regime above \SI{1}{\milli\hertz}, \gls{ttl} coupling is expected to be a major noise contribution after \gls{tdi} \cite{colpi2024lisadefinitionstudyreport,wanner2024depth}. Here we understand \gls{ttl} as the coupling of angular jitters into the longitudinal interferometric measurement. This can have two components \cite{hartig2025tilt}: geometric \gls{ttl} \cite{hartig2022geometric}, describing changes of the propagation time of the beam, and non-geometric \gls{ttl} \cite{hartig2023non}, related to beam or detector properties. The model used here is linear and can schematically be written as $\sum_i C_i^\text{TTL}\cdot \alpha_i$, with some angle measurements $\alpha_i$. If the coupling coefficients $C^\text{TTL}_i$ are known, they can be partially compensated by using a setup of parallel optical plates called the \gls{bam} \cite{wanner2024depth}. Another mitigation strategy is modeling the effect and then subtracting it as before or during global fit, when moving from the raw data to the higher-level scientific output. 

The rotational \glspl{dof} of the \gls{sc} and \gls{mosa} can be measured via \gls{dws} \cite{armano_sensor_2022}. The mechanism determines the pitch and yaw angles between a tilted measurement and reference beam. When we then compare this \gls{dws} output to the longitudinal measurements, we can estimate the \gls{ttl} coupling coefficients.

The impact of \gls{ttl} on the interferometric output has been studied extensively in different settings: focusing on instrumental noise \cite{wanner2024depth,hartig2025tilt,george2023fisher,paczkowski_postprocessing_2022,houba2023time,wang2025postprocessing,chen2025characteristics}, with gravitational wave contributions \cite{hartig2025postprocessing,houba_maneuver_gw_22}, and with intentional excitations of the rotational \gls{dof} called maneuvers \cite{wegener2025design,houba2022optimal,houba_maneuver_gw_22}. Different \gls{tdi} algorithms can be used as well, such as the Michelson variables \cite{wanner2024depth,hartig2025tilt,george2023fisher,wegener2025design,hartig2025postprocessing,paczkowski_postprocessing_2022,houba2022optimal}, null channels \cite{wang2025postprocessing,fang2025tilttolengthnoisesubtractionpointing}, \gls{tdi}-$\infty$ \cite{houba2023time}, or comparative approaches \cite{chen2025characteristics}. The parameter can be inferred using an \gls{ls} estimator (with variations) \cite{hartig2025tilt,wegener2025design,george2023fisher,houba2023time,chen2025characteristics}, or with Bayesian techniques like \gls{mcmc} \cite{hartig2025postprocessing,paczkowski_postprocessing_2022}. For the TianQin mission, \gls{ttl} coupling inference using \gls{tdi} null channels has been studied in a closed-loop dynamics setting \cite{fang2025tilttolengthnoisesubtractionpointing}.

The focus of this work lies in using the full closed-loop dynamics simulation with \gls{dfacs} for the first time to obtain realistic jitter of the rotational \gls{dof} in \gls{lisa} to  model the expected \gls{ttl} coupling contribution in more detail. Parameter inference is implemented via a \gls{ls} estimator, as this is easily interpretable and has a closed-form solution. The accuracy of this estimation is checked in a setting with and without maneuvers.

Previous work on the dynamics modeling include simulations run as predictions for \gls{lisa} \cite{lisa_dfacs_scheme,vidano_lisa_2020,inchauspe23_dynamics,vidano_dfacs_robust_25}, but also analyzing the data obtained from \gls{lisa} Pathfinder \cite{armano_calibrating_2018,lpf_platform_stability}.

The paper is structured as follows: in Sec.~\myhyperref{sec:methods-dynamics} we will introduce the necessary details on the implemented dynamics model. This includes the modeling of the sensing output in Sec.~\myhyperref{sec:methods-sensing} and the control mechanism in Sec.~\myhyperref{sec:methods-control}. Then Secs.~\myhyperref{sec:methods-tdi},\myhyperref{sec:methods-ttl} introduce the description of \gls{tdi} and \gls{ttl}. Lastly, the \gls{ls} estimator is defined in Sec.~\myhyperref{sec:methods-inference} with a possible regularization term.

After this, Sec.~\myhyperref{sec:results-jitter} shows the results on the jitter and \gls{dws} outputs, together with the full interferometric output after \gls{tdi}. The resulting estimated parameters are presented in Sec.~\myhyperref{sec:results-inference}, with several tests on the dependence on the coupling coefficient level. Then the first experiment in Sec.~\myhyperref{sec:exp-1} deals with the dependence on \gls{dws} readout noise. In the second experiment in Sec.~\myhyperref{sec:exp-2-man}, maneuvers are defined in the closed-loop context, and their impact on the inference is discussed. Impacts of random \gls{ttl} coefficients are then investigated in the third experiment in Sec.~\myhyperref{sec:exp-3}. After the summary, the appendix gives details on computations of the bias of the estimator and on correlations of the estimated parameters.

\begin{figure}[!ht]
\includegraphics[width=.95\linewidth]{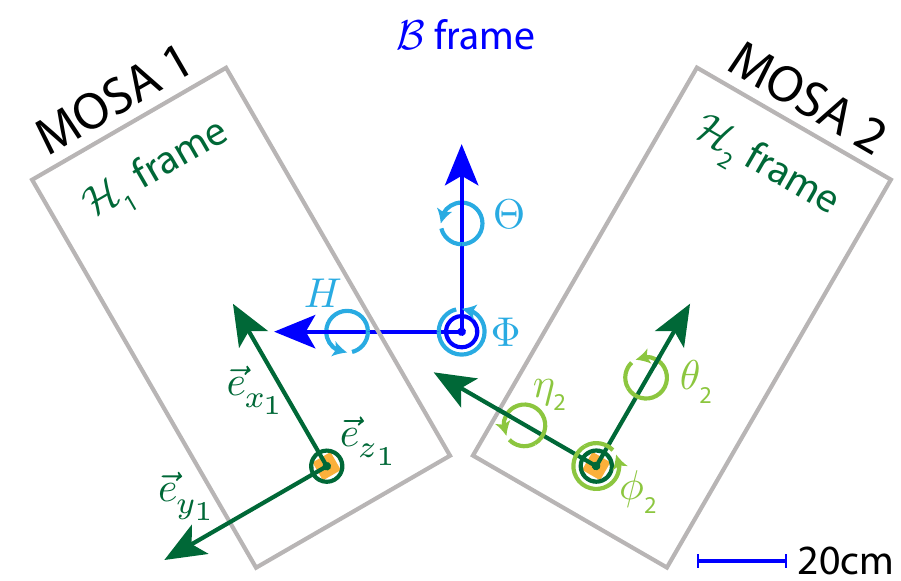}
\caption{Figure cut of \gls{sc} geometry. The definition of the $\mathcal{B}$, $\mathcal{H}_{1}$, and $\mathcal{H}_2$ frames is illustrated. The center-of-mass of the satellite forms the origin of the $\mathcal{B}$ frame, while the centers of the \gls{tm} housings form the origin of the $\mathcal{H}_1$ and $\mathcal{H}_2$ frames. Additionally to the basis vectors, the Cardan angles are defined, following the right-hand rule for their direction of rotation.}
\label{fig:sc}
\end{figure}

As a last part of this section, a small note on notation: the \gls{mosa} are numbered in the \lq scheme\rq\ $\{12,13,23,21,31,32\}$, but especially in the dynamics context, the spacecraft can be regarded as independent as they are not directly connected. Thus sometimes, the left \gls{mosa} of a \gls{sc} will be numbered as $1$, and the right one with $2$ (c.f. Fig.~\myhyperref{fig:sc}). Globally, the left \glspl{mosa} are $\{12,23,31\}$. Vectors and matrices are denoted using bold symbols ($\myvec{\alpha}$, $\myvec{v}$, $\myvec{C}$), and reference frames using a calligraphic font ($\mathcal{B}$, $\mathcal{H}$).

\section{Methods}\label{sec:methods}
\subsection{Dynamics} \label{sec:methods-dynamics}
The full non-linear, time-varying dynamics of the different reference systems of the satellites has been described in a previous publication \cite{inchauspe23_dynamics}. Here we repeat the relevant descriptions for jitters that contribute directly to \gls{ttl} coupling. 

The important reference frames for a specific \gls{sc} are:
\begin{itemize}
    \item $\mathcal{J}$: Galilean (inertial) frame, fixed with respect to distant stars, and with orientation defined according to the International Celestial Reference System convention \cite{kaplan2006iau}.
    \item $\mathcal{O}=\mathcal{B}^*$: defined as the target attitude for the \gls{sc} defined through the precomputed orbits. The origin also follows the ideal orbit of the \gls{sc}.
    \item $\mathcal{B}$: rigidly attached to the \gls{sc} and describing its attitude. The origin is chosen as the center of mass $B$ of the \gls{sc}. Deviations in attitude between $\mathcal{O}$ and $\mathcal{B}$ are the \gls{sc} jitter.
    \item $\mathcal{H}^*_1$, $\mathcal{H}^*_2$: define the target attitude of the \glspl{mosa} with respect to $\mathcal{J}$ frame, based on the incoming laser beam from distant \gls{sc}. The origin is the center of the respective \gls{tm} housing frames.
    \item $\mathcal{H}_1$, $\mathcal{H}_2$: rigidly attached to their respective \gls{mosa} and are defined by their actual attitude inside the \gls{sc}. The origin is the pivot point of the \gls{mosa} rotation, which coincides with the center of the \gls{tm} housing frames $H_1$, $H_2$. Deviations in attitude between $\mathcal{B}$ and $\mathcal{H}_i$ are the \gls{mosa} jitter.
\end{itemize}
The $\mathcal{B}$ and $\mathcal{H}$ frames are also visualized in Fig.~\myhyperref{fig:sc}, showing a cut through the \gls{sc} with the dimensions used in this simulation. The basis vectors for the references frames are illustrated, with the corresponding Cardan angles. The Cardan angles are a specific choice of Euler angles with the convention ZYX/1-2-3, i.e., in a transformation, the $\Phi$ rotation around $z$ is performed first, then $H$ around $y$, and at last $\Theta$ around $x$ \cite{diebel_representing_2006}. The direction of rotation is chosen to be compatible with usual right-hand-rule.

The dynamics of the satellites is largely independent of each other. Thus the \gls{eom} can be formulated for each \gls{sc} separately. We will give here again only the necessary equations that are directly relevant for \gls{ttl}. A full description with derivation can be found in the previous publication \cite{inchauspe23_dynamics}. 

For the \gls{sc} attitude, the interest is in the evolution of $\myexprlower{\myvec{\alpha}}{\mathcal{B}/\mathcal{O}}=(\scalwithref{\theta}{B}{O},\scalwithref{\eta}{B}{O},\scalwithref{\phi}{B}{O})^T$. The associated angular velocity $\vecwithrefrel{\omega}{B}{O}{B}$ given w.r.t. $\mathcal{B}$ frame is related non-trivially with the derivative of the attitude angles \cite{diebel_representing_2006}, via
\begin{equation}
    \begin{split}
        \vecwithrefrel{\myvec{\omega}}{B}{O}{B} &= \bm{E} \left( \myexprlower{\myvec{\alpha}}{\mathcal{B}/\mathcal{O}} \right) \frac{\derd \myexprlower{\myvec{\alpha}}{\mathcal{B}/\mathcal{O}} }{\derd t} \\
        &\coloneqq
        \begin{pmatrix}
            & 1 & 0 & -\sint{H}\\
            & 0 & \cost{\Theta} & \cost{H} \sint{\Theta} \\
            & 0 & -\sint{\Theta} & \cost{H} \cost{\Theta}
        \end{pmatrix}
        \begin{pmatrix}
            \dot{\Theta} \\
            \dot{H} \\
            \dot{\Phi}
        \end{pmatrix}\ ,
    \end{split}
\end{equation}
with the following shorthand $\Theta\coloneqq\scalwithref{\theta}{B}{O}$, $H\coloneqq\scalwithref{\eta}{B}{O}$, $\Phi\coloneqq\scalwithref{\phi}{B}{O}$. 

The \gls{eom} for the \gls{sc} attitude can be formulated as 
\begin{equation}
    \begin{split}
        & \momentinertia{sc}{B}{B} \vecwithrefrel{\dot{\omega}}{B}{O}{B} + \momentinertia{sc}{B}{B} \left[ \trafo{B}{O} \trafo{O}{J} \vecwithrefrel{\omega}{O}{J}{J}\right] \times \vecwithrefrel{\omega}{B}{O}{B} \\
        & + \vecwithrefrel{\omega}{B}{O}{B} \times \left[ \momentinertia{sc}{B}{B} \vecwithrefrel{\omega}{B}{O}{B} \right]\\
        &- \left[\momentinertia{sc}{B}{B} \trafo{B}{O} \trafo{O}{J} \vecwithrefrel{\omega}{O}{J}{J} \right] \times \vecwithrefrel{\omega}{B}{O}{B} \\
        & + \left[ \trafo{B}{O} \trafo{O}{J} \vecwithrefrel{\omega}{O}{J}{J} \right] \times \momentinertia{sc}{B}{B} \vecwithrefrel{\omega}{B}{O}{B}
            \\
        & = \sum \myvec{t}_B^\mathcal{B} - \momentinertia{sc}{B}{B} \trafo{B}{O} \trafo{O}{J} \vecwithrefrel{\dot{\omega}}{O}{J}{J} \\
        &\ \ \ \ - \left[ \trafo{B}{O} \trafo{O}{J} \vecwithrefrel{\omega}{O}{J}{J} \right]  \times \left[\momentinertia{sc}{B}{B} \trafo{B}{O} \trafo{O}{J} \vecwithrefrel{\omega}{O}{J}{J} \right]\, .
    \end{split}
    \label{eq:eom-sc-attitude}
\end{equation}
The matrices $\trafo{B}{O}$ and $\trafo{O}{J}$ are transformation matrices between reference frames. For example, $\trafo{B}{O}$ is the transformation from $\mathcal{O}$ to $\mathcal{B}$ frame. The moment of inertia of the \gls{sc} inside its frame is denoted by $\momentinertia{sc}{B}{B}$. On the right hand side of the equation are all the non-dynamical contributions: externally applied torques $\sum \myvec{t}_B^\mathcal{B}$ (from environment or control systems), and contributions from the orbital motion of the non-inertial reference frame $\mathcal{O}$. Note that $\vecwithrefrel{\omega}{O}{J}{J}$ and $\vecwithrefrel{\dot{\omega}}{O}{J}{J}$ are not dynamical and can be calculated directly from the pre-computed orbits \cite{inchauspe23_dynamics}.

The next important equations are for the attitudes of the \glspl{mosa}. The quantities of interest are the Cardan angle vectors $\myexprlower{\myvec{\alpha}}{\mathcal{H}_\myindex{1}/\mathcal{B}}$, $\myexprlower{\myvec{\alpha}}{\mathcal{H}_\myindex{2}/\mathcal{B}}$. Again, their derivatives are non-trivially related to the angular velocities via 
\begin{equation}
    \begin{split}
        \vecwithrefrel{\omega}{H}[i]{B}{H}[i] &= E \left( \myexprlower{\myvec{\alpha}}{\mathcal{H}_\myindex{i}/\mathcal{B}} \right) \frac{\derd \myexprlower{\myvec{\alpha}}{\mathcal{H}_\myindex{i}/\mathcal{B}} }{\derd t} \ .
    \end{split}
\end{equation}
The \gls{eom} for the \gls{mosa} attitude are given by
\begin{equation}
    \begin{split}
        & \momentinertia{mo}{Q}[i]{H}[i] \vecwithrefrel{\dot{\omega}}{H}[i]{B}{H}[i] + \vecwithrefrel{\omega}{H}[i]{B}{H}[i] \times \left[\momentinertia{mo}{Q}[i]{H}[i] \vecwithrefrel{\omega}{H}[i]{B}{H}[i]\right] \\
        &+ \left[ \trafo{H}[i]{B} \vecwithrefrel{\omega}{B}{O}{B} \right] \times \left[\momentinertia{mo}{Q}[i]{H}[i] \vecwithrefrel{\omega}{H}[i]{B}{H}[i]\right] \\
        &+ \left[ \trafo{H}[i]{B} \trafo{B}{O} \trafo{O}{J} \vecwithrefrel{\omega}{O}{J}{J} \right] \times \left[\momentinertia{mo}{Q}[i]{H}[i] \vecwithrefrel{\omega}{H}[i]{B}{H}[i]\right] \\
        &= \sum \myvec{t}_{H_\myindex{i}}^{\text{rel,} \mathcal{H}_\myindex{i}}\ ,
    \end{split}
    \label{eq:eom-mosa-attitude}
\end{equation}
for $i\in\{1,2\}$. Here $\momentinertia{mo}{Q}[i]{H}[i]$ is the moment of inertia of the \gls{mosa} with respect to its center of mass $Q_i$, and $\sum \myvec{t}_{H_\myindex{i}}^{\text{rel,} \mathcal{H}_\myindex{i}}$ the sum of externally applied torques.

In this publication we focus on this non-linear description, and do not use the linearized and time-averaged system also derived in \cite{inchauspe23_dynamics}. This is because we want to include effects like changes in the \gls{mosa} opening angle in the constellation. Also, the linearization makes the \gls{mosa} \gls{dof} non-dynamical, which we do not want to assume here. Furthermore, we extended the description in the previous publication slightly, by not forcing the angular jitter of the \glspl{mosa} to be zero for $\eta$, as we are later interested in this contribution. The equations in the previous publication also hold for this more general case. For a more realistic modeling, additional stiffness terms would need to be included that represent that while $\phi$ rotations are possible for the \gls{mosa}, movement in $\theta$ and $\eta$ should be heavily suppressed. We neglect this here as the contribution of $\eta$ jitter of the \glspl{mosa} is already very subdominant in the simulation. In other publications that use it for \gls{ttl}, it is usually taken to be a certain fraction of the $\phi$ jitter of the \gls{mosa} \cite{wanner2024depth,george2023fisher,wegener2025design}; it is even set to zero in \cite{hartig2025tilt}.

\subsection{Sensing}\label{sec:methods-sensing}
Here we present the implemented sensing system relevant for \gls{ttl}, the \acrfull{dws} for the inter-spacecraft interferometer. This is an actual observable of the simulation, in contrast with the internal states of the simulation introduced in the previous section on dynamics.

The \gls{dws} is used to determine the pitch ($\eta$) and yaw ($\phi$) of the \gls{mosa} with respect to an incoming beam. Using the reference frames defined in the previous subsection, the incoming beam is just along $\framevu{H^*}{x}$. Transforming the actual direction of the \gls{mosa} in $\mathcal{H}$ frame (trivially $\framevu{H}{x}$) to $\mathcal{H}^*$ frame, introduce the following sequence of rotations that include breathing angles as well as \gls{sc} and \gls{mosa} jitter. In formulas, this is given by
\begin{align}
    \mymat{T}_i \coloneqq \mymat{T}_{\mathcal{O}}^{\mathcal{H}^*_\myindex{i}} \trafo{O}{B}\trafo{B}{H}[i]\ ,
\end{align}
which can be written out with standard rotation matrices around $x$, $y$, and $z$ axes, taking care of the correct ordering
\begin{align}
    \trafomat{z}(\varphi_b^\pm)\left[\trafomat{x}(\Theta) \trafomat{y}(H) \trafomat{z}(\Phi)\right]^T\! \left[\trafomat{y}(\eta_i) \trafomat{z}(\varphi_b^\pm+\phi_i)\right]^T
\end{align}
with the \gls{mosa} angles $\eta_i \equiv \scalwithref{\eta}{H}[i]{B}$, $\phi_i \equiv \scalwithref{\phi}{H}[i]{B}$, $i\in\{1,2\}$, and the \gls{sc} angles $\Theta \equiv \scalwithref{\Theta}{B}{O}$, $H \equiv \scalwithref{H}{B}{O}$, $\Phi \equiv \scalwithref{\Phi}{B}{O}$. Here the breathing angle $\varphi_b$ has been split into the signed half-angles $\varphi_b^\pm\coloneqq \pm \varphi_b / 2$. Note that passive rotations are used as a convention \cite{inchauspe23_dynamics}. As all angles apart from the telescope opening angles $\varphi_b^\pm$ are small, use the first-order approximations $\cos{x}\approx 1$ and $\sin{x}\approx x$ for the rotation matrices. This yields for $i\in\{1,2\}$
\begin{align}
    \mymat{T}_i=\begin{pmatrix} 1 & -\Phi-\phi_i & \eta_i+ H c - \Theta s \\
    \Phi+\phi_i & 1 & -\Theta c - H s \\ 
    -\eta_i- H c + \Theta s & \Theta c + H s & 1
    \end{pmatrix}
\end{align}
using $c\equiv \cos(\varphi_b^\pm)$ and $s\equiv \sin(\varphi_b^\pm)$. 

Given this transformation between reference frames, the pitch and yaw angles can be read off using \cite{diebel_representing_2006}
\begin{align}
    \eta_i^\text{dws} &= -\operatorname{asin}(\mymat{T}_i\framevu{H}{x}\cdot \framevu{H^*}{z}) + w^\eta_i\ , \label{eq:dws-nonlin-eta}\\
    \phi_i^\text{dws} &= \operatorname{atan2}(\mymat{T}_i\framevu{H}{x}\cdot \framevu{H^*}{y},\mymat{T}_i\framevu{H}{x}\cdot \framevu{H^*}{x}) + w^\phi_i\ . \label{eq:dws-nonlin-phi}
\end{align}
with additional white noise contributions $w^\eta_i$, $w^\phi_i$, $i\in\{1,2\}$, simulating readout noise. Using the same approximation for the sine and cosine functions as before, this gives the following for the \gls{dws} angles
\begin{align}
    \eta_1^\text{dws} &\approx \eta_1 + H \cos(\varphi_b^+) - \Theta \sin(\varphi_b^+) + w^\eta_1 \label{eq:dws-approx-eta1}\\
    \phi_1^\text{dws} &\approx \Phi + \phi_1 + w^\phi_1\label{eq:dws-approx-phi1}\\
    \eta_2^\text{dws} &\approx \eta_2 + H \cos(\varphi_b^-) - \Theta \sin(\varphi_b^-) +  w^\eta_2 \label{eq:dws-approx-eta2}\\
    \phi_2^\text{dws} &\approx \Phi + \phi_2 +  w^\phi_2 \label{eq:dws-approx-phi2}
\end{align}
for the left and right \glspl{mosa}. Further approximations can be made for the pitch angles, setting $\varphi_b^\pm=\pm \pi/6$ to get 
\begin{align}
    \eta_1^\text{dws} &\approx \eta_1 + \frac{\sqrt{3}}{2} H - \frac{1}{2} \Theta +  w^\eta_1\ , \label{eq:dws-approx-eta1-equal}\\
    \eta_2^\text{dws} &\approx \eta_2 + \frac{\sqrt{3}}{2} H + \frac{1}{2} \Theta +  w^\eta_2\ , \label{eq:dws-approx-eta2-equal}
\end{align}
which are the usual modeling outputs in the \gls{ttl} literature \cite{wanner2024depth,george2023fisher,wegener2025design,hartig2025tilt,paczkowski_postprocessing_2022}. In our modeling, only the \gls{dfacs} controller receives this approximation, the rest goes beyond this and directly implements Eqs.~\myhyperref{eq:dws-nonlin-eta},\myhyperref{eq:dws-nonlin-phi}. However, without significant excitations in the dynamical system, the first-order approximation in the jitter angles is very accurate. The deviation in the \gls{mosa} breathing angles $\varphi_b^\pm$ from $\pm \pi/6$ leads to a percent level error, which is further investigated in App.~\myhyperref{app:long-term-var}.

\subsection{Control and Actuation}\label{sec:methods-control}
In this subsection we summarize the aspects of the \gls{dfacs} relevant to \gls{ttl} coupling from \cite{lisa_dfacs_scheme,inchauspe23_dynamics}: \gls{mosa} actuation and \gls{mps} for the \gls{sc} attitude. The \gls{mosa} actuation can only operate along the $\phi$ angle and moves the two \glspl{mosa} on one \gls{sc} symmetrically.

As the controller should only see output from sensing, the \gls{sc} angles are derived from the measured \gls{dws} angles. This is based on \cite{lisa_dfacs_scheme}, but note the sign changes. The recipe is the following
\begin{align}
    \Theta^\text{dws} &\approx \eta_2^\text{dws} - \eta_1^\text{dws}\ , \\
    H^\text{dws} &\approx \frac{1}{\sqrt{3}}\left(\eta_1^\text{dws} + \eta_2^\text{dws} \right) \ ,\\
    \Phi^\text{dws} &\approx \frac{1}{2}\left(\phi_1^\text{dws} + \phi_2^\text{dws} \right) \ .
\end{align}
Using the approximations from the previous section for the \gls{dws} angles, this yields
\begin{align}
    \Theta^\text{dws} &\approx \eta_2 - \eta_1 + \Theta +  w^\eta_2 -  w^\eta_1\ , \label{eq:dws-control-sc-angles1}\\
    H^\text{dws} &\approx \frac{1}{\sqrt{3}}\left(\eta_1 + \eta_2 \right) + H + \frac{1}{\sqrt{3}}\left(w^\eta_1 + w^\eta_2 \right)\ ,\label{eq:dws-control-sc-angles2}\\
    \Phi^\text{dws} &\approx \frac{1}{2}\left(\phi_1 + \phi_2 \right) + \Phi + \frac{1}{2}\left(w^\phi_1 + w^\phi_2 \right)\ . \label{eq:dws-control-sc-angles3}
\end{align}
This also shows that the white noise for the measured \gls{dws} angles contributes with different strength to the inferred \gls{sc} angles: $\Theta$ receives double the noise, $H$ roughly \num{1.15} times the original noise, and $\Phi$ the same as the \gls{dws} angles.

The \gls{mosa} actuation controls $\phi_1^\text{dws} - \phi_2^\text{dws}$. Approximating this gives
\begin{align}
    \phi_1^\text{dws} - \phi_2^\text{dws} \approx \phi_1 - \phi_2 + w^\phi_1 - w^\phi_2 \ ,
\end{align}
which shows that this only contains the \gls{mosa} jitters and the \gls{dws} noise, but no contribution from the \gls{sc} jitters.

For the control system, these four quantities $(\Theta^\text{dws},$ $H^\text{dws},\Phi^\text{dws},\phi_1^\text{dws} - \phi_2^\text{dws})$ are used as an input for a dynamical system, determining the controller output. This is modeled with matrices in a state-space representation. This controller output is then fed into another dynamical system, determining the actuation response to the controller signal.

In the end, the commanded actuation output returns forces and torques that should be applied to the \gls{sc} or the \gls{mosa} opening angles. To make this more realistic, white noise is added to these forces and torques, before these are looped back into the \gls{eom} (c.f. Eqs.~\myhyperref{eq:eom-sc-attitude} and \myhyperref{eq:eom-mosa-attitude}).

\subsection{Closed-Loop Simulation}\label{sec:methods-cls}
The relevant elements of simulation have been shown in the previous sections: the \gls{eom} determine the response of the physical system given the internal states and the external forces/torques. Then the sensing simulates a measurement of some internal states, which are then fed into the controller and actuation, determining the response of the control loop. These forces/torques are then in turn fed into the \gls{eom} as external inputs in the next time step.

It is important to note that the noises added in sensing and actuation do not simply add to the sensing output, but also are a source for jitter via the closed loop. A full list of noise settings can be found in App.~\myhyperref{app:lisanode}.

For the actual simulation of the data, the LISANode suite \cite{bayle_lisanode} is used. Details of the implementation can be found in App.~\myhyperref{app:lisanode}. It is important to note that in order to make the computations more tractable, instead of working with phases as the output of the interferometer, the simulation represents data via frequencies \cite{bayle2023unified}. Concretely, for a given periodic signal $A\cdot \cos(\Phi(t))$ with amplitude $A$ and total phase $\Phi(t)$ in radians, the corresponding angular frequency $\omega(t)$ in \unit[per-mode=symbol]{\radian\per\second} is given by 
\begin{equation}
    \omega(t) = \dot{\Phi}(t)\ ,
\end{equation}
such that the signal has the alternative representation
\begin{align}
    A\cdot\cos\left(\int_0^t \operatorname{d}\!t' \, \omega(t')\right)\ .
\end{align}
Thus, for example, the \gls{dws} output of the simulation $\dot{\eta}_i^\text{dws}$, $\dot{\phi}_i^\text{dws}$ for $i\in\{1,2\}$ will also have units \unit[per-mode=symbol]{\radian\per\second}, with a noise contribution that is no longer white. 

\subsection{Time-Delay Interferometry}\label{sec:methods-tdi}
\Gls{tdi} is used to post-process the single-link measurements on each \gls{sc} into combined data that suppresses the dominant laser frequency noise \cite{tinto_time-delay_2020}. In this work the second-generation Michelson variables \cite{bayle:phd,otto:phd} are used to account for changing arm lengths within the constellation over time. The basis for the formulation are the operators $D_{ij}$ that delay a measurement by the light travel times $\tau_{ij}$ along a certain link. The naming convention can be read off from Fig.~\myhyperref{fig:constellation}. The action of the operator is defined as
\begin{equation}
    D_{ij}(t) \Phi(t) \coloneqq \Phi(t-\tau_{ij}(t)) \ ,
\end{equation}
where $t$ corresponds to a synchronized time coordinate between the \gls{sc} (for example extracted from pseudo-ranging). The delay operator can be concatenated for ease of notation like $D_{ijk}\coloneqq D_{ij}D_{jk}$, which can then be iteratively extended to longer chains.

As mentioned in the last section, we generally work with frequencies instead of phases. This means we also need the derivatives of the time-shifted signal \cite{bayle2021adapting}
\begin{equation}
    \frac{\operatorname{d}}{\operatorname{d}t} D_{ij}(t) \Phi(t) \coloneqq (1-\dot{\tau}_{ij}) D_{ij}(t)\dot{\Phi}(t) \eqqcolon \dot{D}_{ij} \omega(t) \ ,
\end{equation}
causing the frequency signal not only to be shifted in time, but also in amplitude by a Doppler factor. This can then be used to define a new operator $\dot{D}_{ij}$.

Let $\{\bar{\eta}_{ij} | i,j\in\{1,2,3\} \land i \neq j \}$ be a certain combination of beatnote measurements in frequency units, called the \gls{tdi} intermediate variables \cite{paczkowski_postprocessing_2022,bayle2021adapting}. The operator for the TDI-X combination is defined as
\begin{align}
    \begin{split}
    \operatorname{TDI}_X[\bar{\eta}_{ij}] \coloneqq &(1-\dot{D}_{121}-\dot{D}_{12131}+\dot{D}_{1312121}) \\
    &\cdot (\bar{\eta}_{13}+\dot{D}_{13}\bar{\eta}_{31}) \\
    &- (1-\dot{D}_{131}-\dot{D}_{13121}+\dot{D}_{1213131}) \\ 
    &\cdot (\bar{\eta}_{12}+\dot{D}_{12}\bar{\eta}_{21}) \ .
    \end{split}\label{eq:def-tdi-operator}
\end{align}
The operators $\operatorname{TDI}_Y$, $\operatorname{TDI}_Z$ can be found by cyclical permutations of this equation. These combinations can be interpreted as interference of virtual photon paths and were detailed in other publications \cite{tinto_time-delay_2020,martens2021trajectory}.

In terms of software, the PyTDI suite \cite{staab_pytdi} is used for processing the outputs of LISANode.

\subsection{Tilt-To-Length Coupling}\label{sec:methods-ttl}
In interferometry, \gls{ttl} coupling refers to the coupling of motion orthogonal to the beam into the interferometric readout \cite{hartig2025tilt}. We investigate here the dominant contribution \cite{wanner2024depth} due to a coupling to angular jitters in the typical linear approximation, which is schematically given by
\begin{equation}
    \sum_{i} C_i^\text{TTL} \cdot \alpha_i \ ,
\end{equation}
using angle measurements $\alpha_i$. The coupling coefficients $C_i^\text{TTL}$ have units of \unit[per-mode=symbol]{\metre\per\radian}, so that the total contribution has the unit \unit{\metre}. As LISANode works with frequencies instead of phases, this needs to be rewritten as \cite{lisainstrument_doc}
\begin{equation}
    S_\text{TTL} = -\frac{\nu_0}{c}\sum_{i} C_i^\text{TTL} \cdot \dot{\alpha}_i \ ,
\end{equation}
with the \gls{lisa} laser frequency $\nu_0 = 2.816\cdot 10^{14}\,\unit{\hertz}$ and speed of light $c$. The minus sign comes from considerations about the impact on the phase (a positive coefficient corresponds to a phase reduction). The units are then $[\dot{\alpha}_i]=\unit[per-mode=symbol]{\radian\per\second}$ and thus $[S_\text{TTL}]=\unit{\hertz}$.

Following \cite{hartig2025tilt} to arrive at an appropriate frequency version, in \gls{lisa} the \gls{ttl} contribution along a single link (\gls{sc} $i$ receiving light from \gls{sc} $j$) is given by
\begin{align}
\begin{split}
    S_{\text{TTL}}^{ij} =\,& -\frac{\nu_0}{c} \left(C_{ij\phi}^\text{Rx}\cdot \dot{\phi}_{ij} + C_{ij\eta}^\text{Rx}\cdot \dot{\eta}_{ij} \right. \\
    &+ \left. C_{ji\phi}^\text{Tx}\cdot \dot{D}_{ij}\dot{\phi}_{ji} + C_{ji\eta}^\text{Tx}\cdot \dot{D}_{ij}\dot{\eta}_{ji} \right) \\
    \eqqcolon\,& S_{\text{TTL},\eta}^{ij} + S_{\text{TTL},\phi}^{ij}\ ,
\end{split}
\end{align}
where $\dot{\eta}_{ij}$, $\dot{\phi}_{ij}$ are the derivatives of the total \gls{mosa} angles, i.e., the angles measured by the \gls{dws} without the noise (c.f. Eqs.~\myhyperref{eq:dws-approx-phi1}--\myhyperref{eq:dws-approx-eta2-equal}). All together, this yields $4$ parameters for each of the $6$ links, so $24$ parameters determining the \gls{ttl} contribution.

The interferometric readout along a single link is dominated by laser frequency noise, as mentioned in the last section. Thus, the \gls{ttl} contribution to \gls{tdi} needs to be investigated. Using the definition of the \gls{tdi} operators in Eq.~\myhyperref{eq:def-tdi-operator}, the result $S_{\text{TTL},\eta}^\text{TDI-X}$ is
\begin{widetext}
\begin{align}
    \begin{split}
    \operatorname{TDI}_X[S_{\text{TTL},\eta}^{ij}] = &-\frac{\nu_0}{c}\left(1-\dot{D}_{121}-\dot{D}_{12131}-\dot{D}_{1312121}\right) \left(\left(C_{13\eta}^\text{Rx}+C_{13\eta}^\text{Tx}\dot{D}_{131}\right)\dot{\eta}_{13} + \left(C_{31\eta}^\text{Rx}+C_{31\eta}^\text{Tx}\right)\dot{D}_{13}\dot{\eta}_{31}\right) \\
    &+\frac{\nu_0}{c} \left(1-\dot{D}_{131}-\dot{D}_{13121}-\dot{D}_{1213131}\right) \left(\left(C_{12\eta}^\text{Rx}+C_{12\eta}^\text{Tx}\dot{D}_{121}\right)\dot{\eta}_{12} + \left(C_{21\eta}^\text{Rx}+C_{21\eta}^\text{Tx}\right)\dot{D}_{12}\dot{\eta}_{21}\right)
    \end{split}\label{eq:def-tdix-ttl}
\end{align}
\end{widetext}
with $S_{\text{TTL},\eta}^\text{TDI-Y}$ and $S_{\text{TTL},\eta}^\text{TDI-Z}$ again given by cyclical permutations. The definition is then identical for the $\phi$ contribution. In total we have
\begin{align}
    S_{\text{TTL}}^\text{TDI-X} &= S_{\text{TTL},\eta}^\text{TDI-X} + S_{\text{TTL},\phi}^\text{TDI-X}\ , \\
    S_{\text{TTL}}^\text{TDI} &= S_{\text{TTL}}^\text{TDI-X} + S_{\text{TTL}}^\text{TDI-Y} + S_{\text{TTL}}^\text{TDI-Z}\ .
\end{align}
These can then be split into contributions from the $12$ total angular velocity measurements $\dot{\alpha}_{ij}$ ($2$ angles of $2$ \glspl{mosa} on $3$ \gls{sc}) and their $12$ delayed counterparts $\dot{D}_{ji}\dot{\alpha}_{ij}$. This is detailed in the appendix of \cite{wegener2025design}.

For a compact notation, let $\dot{\myvec{\alpha}}_\text{TDI-X}\in\mathbb{R}^{1,24}$ be the collection of the $24$ angular velocities after the \gls{tdi}-X algorithm, arranged in the following way
\begin{align}
\begin{split}
    \dot{\myvec{\alpha}}_\text{TDI-X} = &-\frac{\nu_0}{c}\left(\operatorname{TDI}_X[\dot{\eta}_{12}] \,\cdots\, \operatorname{TDI}_X[\dot{\phi}_{12}]\,\cdots\right. \\
    &\left.\ \operatorname{TDI}_X[\dot{D}_{21}\dot{\eta}_{12}]\,\cdots\,
    \operatorname{TDI}_X[\dot{D}_{21}\dot{\phi}_{12}]\,\cdots
    \right)
\end{split}
\end{align}
with the order of the links fixed to $\{12,13,23,21,31,32\}$. Define analogously for \gls{tdi}-Y and \gls{tdi}-Z. These can then in turn be combined into $\dot{\mymat{\alpha}}_\text{TDI}\in\mathbb{R}^{3,24}$. For the \gls{dws} readout, define $\dot{\mymat{\alpha}}_\text{TDI}^\text{dws}\in\mathbb{R}^{3,24}$ in a similar way, by replacing the total \gls{mosa} angles $\dot{\eta}_{ij}$, $\dot{\phi}_{ij}$ by the \gls{dws} measurements $\dot{\eta}_{ij}^\text{dws}$, $\dot{\phi}_{ij}^\text{dws}$. The units are $[\dot{\mymat{\alpha}}_\text{TDI}] = \unit[per-mode=symbol]{\Hz\radian\per\m}$.

Lastly, also arrange the full interferometric readout $S_\text{IFO}^{ij}$ in a vector $\myvec{S}_\text{IFO}^\text{TDI}\in\mathbb{R}^{3,1}$
\begin{equation}
    \myvec{S}_\text{IFO}^\text{TDI} = \begin{pmatrix}
        \operatorname{TDI}_X[S_\text{IFO}^{ij}] \\
        \operatorname{TDI}_Y[S_\text{IFO}^{ij}] \\
        \operatorname{TDI}_Z[S_\text{IFO}^{ij}]
    \end{pmatrix}\ ,
\end{equation}
as well as the \gls{ttl} parameters in a vector $\myvec{C}_\text{TTL}\in\mathbb{R}^{24,1}$
\begin{equation}
    \myvec{C}_\text{TTL} = \left(C_{12\eta}^\text{Rx} \,\cdots\, C_{12\phi}^\text{Rx}\,\cdots\, C_{12\eta}^\text{Tx} \,\cdots\, C_{12\phi}^\text{Tx} \cdots\right)^T\ .
\end{equation}
This then for example enables to write $S_{\text{TTL}}^\text{TDI}$ in the more compact form $\sum_{i=1}^3 \left(\dot{\mymat{\alpha}}_\text{TDI}\cdot \myvec{C}_\text{TTL}\right)_i$.

Note that while the introduced notation here refers only to single variables, this can be directly extended to time series, which we will often be working with. In this case, $\myvec{S}_\text{IFO}^\text{TDI}$ becomes a vector in $\mathbb{R}^{3N,1}$, and $\dot{\mymat{\alpha}}_\text{TDI},\dot{\mymat{\alpha}}_\text{TDI}^\text{dws}\in\mathbb{R}^{3N,24}$. 

\subsection{Parameter Estimation}\label{sec:methods-inference}
A variety of algorithms are available to infer parameters from data. Commonly used in the \gls{ttl} coupling literature are the \acrfull{ls} and Monte-Carlo-Markov-Chain techniques. Here we will focus on the simpler \gls{ls} estimator, as it is well interpretable and has a closed-form solution.

The inference problem is phrased as a minimization problem
\begin{equation}
    \hat{\myvec{C}}_\text{TTL} = \underset{\myvec{C}\in\mathbb{R}^{24}}{\operatorname{argmin}}\left\lVert \myvec{S}_\text{IFO}^\text{TDI} + \dot{\myvec{\alpha}}_\text{TDI}^\text{dws}\cdot \myvec{C}\right\rVert_2^2 \ , \label{eq:cost-ls}
\end{equation}
where the subscript 2 specifies the Euclidean norm and the plus sign is used instead of the usual minus as we are working with frequencies, not phases. This function has a global minimum, which can be found directly by differentiation. The unique solution to the problem is then
\begin{equation}
    \hat{\myvec{C}}_\text{TTL} = -\left(({{\dot{\myvec{\alpha}}}_\text{TDI}^{\text{dws}}})^T \cdot {\dot{\myvec{\alpha}}}_\text{TDI}^\text{dws} \right)^{-1} ({{\dot{\myvec{\alpha}}}_\text{TDI}^{\text{dws}}})^T \cdot \myvec{S}_\text{IFO}^\text{TDI}\ . \label{eq:ls-solution}
\end{equation}
Unfortunately, it has already been shown that this estimator is biased, i.e., 
\begin{equation}
    \lim_{N\rightarrow\infty} \hat{\myvec{C}}_\text{TTL} - \myvec{C}_\text{TTL} \neq 0
\end{equation}
for non-white jitter shapes \cite{hartig2025tilt}. The explicit formulas for the bias have been extended for non-white \gls{dws} readout noise in App.~\myhyperref{app:bias}. In \cite{hartig2025tilt}, the proposed solution is using an Instrumental Variables simulator, closely related to \gls{ls}. However, this does not work here as a crucial prerequisite is that the \gls{dws} readout noise is white.

If the eigenvalues of $({\dot{\myvec{\alpha}}_\text{TDI}^{\text{dws}}})^T \cdot \dot{\myvec{\alpha}}_\text{TDI}^\text{dws}$ get close to zero, the inference problem becomes ill-conditioned. This can be regularized, with a simple choice being the Tikhonov regularization \cite{regularization_inverse_problems}. This has the advantage that the global minimum can still be calculated analytically. In this regularization, add a term that aims to keep the coefficients small
\begin{equation}
    \underset{\myvec{C}\in\mathbb{R}^{24}}{\operatorname{argmin}}\left(\left\lVert \myvec{S}_\text{IFO}^\text{TDI} + \dot{\myvec{\alpha}}_\text{TDI}^\text{dws}\cdot \myvec{C} \right\rVert_2^2 + 3N\lambda \left\lVert \myvec{C} \right\rVert_2^2 \right)
\end{equation}
for a given $\lambda\in\mathbb{R}^+_0$ and $N$ the number of elements in the time series. The factor $3N$ is a customary normalization, as the usual \gls{ls} cost function comes with a $1/(3N)$ prefactor. The solution is invariant under rescalings of the cost function, thus the prefactor is only introduced now. The solution is given by
\begin{align}
    \begin{split}
    \hat{\myvec{C}}_\text{TTL}[\lambda] = &-\left(({\dot{\myvec{\alpha}}_\text{TDI}^{\text{dws}}})^T \dot{\myvec{\alpha}}_\text{TDI}^\text{dws} + 3N\lambda \mathds{1} \right)^{-1}  \\
    &\cdot ({\dot{\myvec{\alpha}}_\text{TDI}^{\text{dws}}})^T \myvec{S}_\text{IFO}^\text{TDI} 
    \end{split}\label{eq:ls-reg-solution}
\end{align}
\begin{figure*}[!ht]
\centerline{\includegraphics[width=\linewidth, clip]{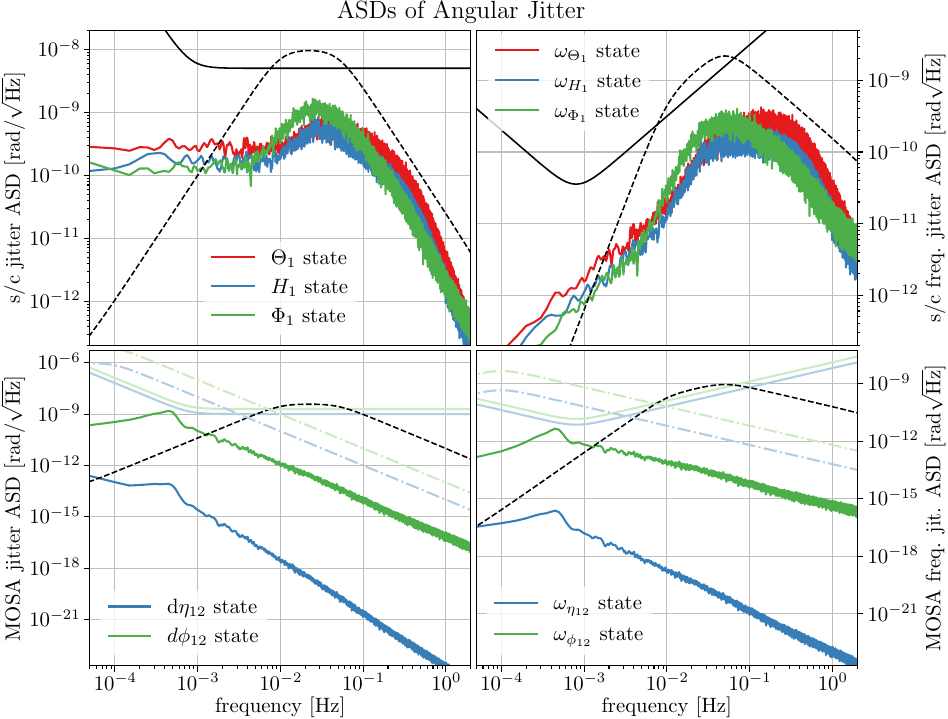}}
\caption{Representative \gls{sc} and \gls{mosa} jitter shapes are shown. The left subplots show the \glspl{asd} of the internal states of the simulator in terms of radians. The right subplots show the later used time derivatives. The \glspl{asd} are calculated from a \SI{e5}{\s} dataset with $N_\text{avg.}=5$ (c.f. App.~\myhyperref{app:lisanode}). Reference lines show jitters used in another publication \cite{hartig2025tilt}: solid lines in black or lighter colors are white jitter references with a low frequency roll off, while the dashed black lines are colored jitter. Note that for the colored jitter case, \gls{mosa} $\eta$ is set to zero. Additionally, for the \gls{mosa} jitter the dash-dotted reference lines have been added from case B in \cite{george2023fisher}. Their \glspl{asd} are defined in Eqs.~\myhyperref{eq:ref_white_sc_jitter}--\myhyperref{eq:ref_col_mosa_jitter2}.\\
The top left subplot has a plateau for low frequencies, which exactly corresponds to the white \gls{dws} readout noise (\SI[power-half-as-sqrt,per-mode=symbol]{0.2}{\nano\radian\per\Hz\tothe{0.5}}). The fall-off towards higher frequencies is a result of the moment-of-inertia. Without \gls{dws} noise, the fall-off would be a power law; but with the noise present, there is a change in the fall-off. The \gls{sc} thruster noise widens the peak.}
\label{fig:jitters-state}
\end{figure*}
where it becomes evident that $\lambda$ has the interpretation of a minimum eigenvalue of the matrix. For $\lambda=0$ the \gls{ls} is recovered. Algorithms exist to compute the parameter $\lambda$, but we will rather use a heuristic ansatz later as these did not give usable solutions.

How to estimate the error of inferred parameters is detailed in App.~\myhyperref{app:error}. Knowing the true parameters, the overall deviation of the inferred parameters is measured with the \gls{rms}, which is defined as for a vector $\myvec{v}\in\mathbb{R}^n$ as
\begin{equation}
    \operatorname{RMS}(\myvec{v}) = \sqrt{\frac{1}{n}\sum_{i=1}^n \myvec{v}_i^2}\ .
\end{equation}
To measure the deviation of just the $\eta$ and $\phi$ parameters, just set the components in the vector $\hat{\myvec{C}}_\text{TTL}-\myvec{C}_\text{TTL}$ that correspond to the other parameters to zero.

As we work here with frequencies instead of phases, the noise in the \gls{dws} output is no longer white, but the derivative of a white noise. Thus, along with the jitters, noises need to be considered colored. Furthermore, the \gls{tdi} channels X, Y, Z are not independent, as would ideally be the case for the \gls{ls} estimator. This can be seen by the fact that \lq orthogonal\rq\ channels A, E, T can be constructed, where only two of them have a large \gls{gw} contribution \cite{lisa_opt_sensitivity_aet_02}.
\begin{figure*}[!ht]
\centerline{\includegraphics[width=\linewidth, clip]{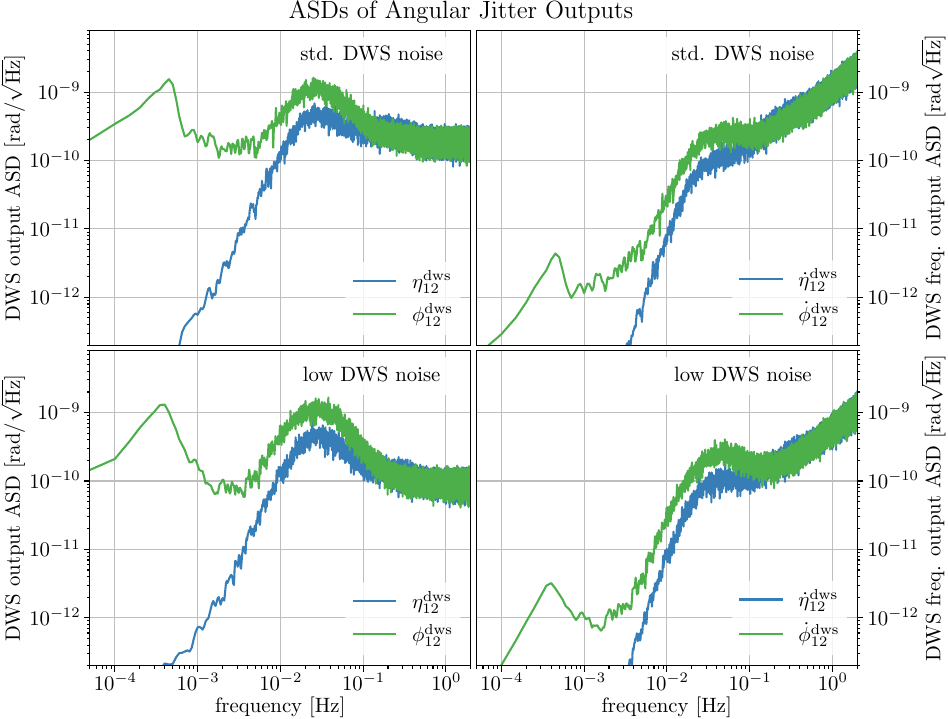}}
\caption{Representative \gls{dws} outputs for \gls{mosa} 12. The left subplots show the \glspl{asd} of the output in terms of radians over time, the right subplots show the later used derivatives. The upper subplots show the result of a simulation with the standard \gls{dws} noise settings of \SI[power-half-as-sqrt,per-mode=symbol]{0.2}{\nano\radian\per\Hz\tothe{0.5}}, and the lower subplots with half the noise. The \glspl{asd} are calculated from a \SI{e5}{\s} dataset with $N_\text{avg.}=5$ (c.f. App.~\myhyperref{app:lisanode}).\\
The left subplots show two peaks for the $\phi_{12}$ sensing output. The low-frequency peak comes from the contribution of the \gls{mosa} jitter, and the medium-frequency peak from the contribution of the \gls{sc} jitter. The $\eta_{12}$ sensing output only sees the \gls{sc} jitter contributions, as the \gls{mosa} jitter is subdominant, resulting in only one peak. The plateau towards higher frequencies corresponds directly to the level of \gls{dws} readout noise.}
\label{fig:jitters-sensing}
\end{figure*}

However, the form of Eq.~\myhyperref{eq:cost-ls} also reveals that the cost function, and thus its solution is invariant under orthogonal transformations of the \gls{tdi} channels. For example, consider an orthogonal matrix that maps the X, Y Z channels to A, E, T. That such a matrix exists can be seen from the defining equations of the orthogonal combinations \cite{lisa_opt_sensitivity_aet_02}. The following will however be true for any orthogonal matrix $\mymat{T}\in\mathbb{R}^{3\times3}$. Given any parameter vector $\myvec{C}\in\mathbb{R}^{24}$ it holds that
\begin{align}
	\begin{split}
	&\left\lVert \myvec{S}_\text{IFO}^\text{TDI} + \dot{\myvec{\alpha}}_\text{TDI}^\text{dws}\cdot \myvec{C}\right\rVert_2^2 \\
    &= \left( \myvec{S}_\text{IFO}^\text{TDI} + \dot{\myvec{\alpha}}_\text{TDI}^\text{dws}\cdot \myvec{C}\right)^T \left( \myvec{S}_\text{IFO}^\text{TDI} + \dot{\myvec{\alpha}}_\text{TDI}^\text{dws}\cdot \myvec{C}\right) \\ 
	&= \left( \myvec{S}_\text{IFO}^\text{TDI} + \dot{\myvec{\alpha}}_\text{TDI}^\text{dws}\cdot \myvec{C}\right)^T \mymat{T}^T \mymat{T} \left( \myvec{S}_\text{IFO}^\text{TDI} + \dot{\myvec{\alpha}}_\text{TDI}^\text{dws}\cdot \myvec{C}\right) \\
	&=\left(\mymat{T} \myvec{S}_\text{IFO}^\text{TDI} + \mymat{T} \dot{\myvec{\alpha}}_\text{TDI}^\text{dws}\cdot \myvec{C}\right)^T \!\!\left( \mymat{T}\myvec{S}_\text{IFO}^\text{TDI} + \mymat{T} \dot{\myvec{\alpha}}_\text{TDI}^\text{dws}\cdot \myvec{C}\right)  \\
	& \left\lVert \myvec{S}_\text{IFO}^\text{TDI*} + \dot{\myvec{\alpha}}_\text{TDI*}^\text{dws}\cdot \myvec{C}\right\rVert_2^2
	\end{split}
\end{align}
where $ \myvec{S}_\text{IFO}^\text{TDI*}$ refers to the new \gls{tdi} channels after the orthogonal transformation and identically for $\dot{\myvec{\alpha}}_\text{TDI*}^\text{dws}$. Note that this applies also when time series are used to describe the \gls{tdi} channels.

Because the inference works in the time-domain, the data will need to be filtered to remove noises that are especially dominant at high frequencies. This does not change the formulas shown in this section beyond having to exchange $\myvec{S}_\text{IFO}^\text{TDI}$ and $\dot{\myvec{\alpha}}_\text{IFO}^\text{TDI}$ with their filtered counterparts. The frequencies for the bandpass filter are \SI{15}{\milli\Hz} to \SI{70}{\milli\Hz} (c.f. App.~\myhyperref{app:filtering}).

\section{Results}
This section presents the central results of this work: the \gls{sc} and \gls{mosa} jitter outputs of the closed-loop dynamics simulation of \gls{lisa}, the resulting estimate of \gls{ttl} coupling in \gls{tdi}, and its implications for the inference of \gls{ttl} parameters in different scenarios. 

\subsection{Jitter Shapes}\label{sec:results-jitter}
In terms of jitters, there are two main outputs that are relevant: the measured \gls{dws} output, and the internal dynamical states of the simulator, which are not available in a non-simulation setup. The results presented here show the jitters of \gls{sc} $1$, and \gls{mosa} $12$ without restricting generality. There are only minor differences in the jitters due to different noise realizations and the choice of primary laser in the locking scheme \cite{fplan_lisa_24}. Furthermore, the simulation and inference are based on changes in frequency, not phase, of the laser beam (c.f. Sec.~\myhyperref{sec:methods-cls}). Thus angular velocities (rad/s) instead of angles (rad) are considered. For better comparison to previous work, the jitter \glspl{asd} also show the output in terms of radian.

Figure~\myhyperref{fig:jitters-state} shows the internal state output for the \gls{sc} and \gls{mosa} jitter. The left column shows this for the angle time series, the right in terms of angular velocities. 

Focusing on the \gls{sc} jitters in terms of angles, the plateau for low frequencies is due to \gls{dws} readout noise. The different levels can be directly explained by the different prefactors when deriving \gls{sc} angles from the \gls{dws} output (c.f. Eqs.~\myhyperref{eq:dws-control-sc-angles1}-\myhyperref{eq:dws-control-sc-angles3}). The noise is present in the internal state due to the closed-loop simulation: the controller for \gls{sc} angles sees the \gls{dws} readout noise, and compensates for this at low frequencies; this then removes the readout noise in the \gls{dws} output for low frequencies, but add this into the internal state. Above $\sim$\SI{8}{\milli\Hz}, the jitter increases due to weakening control power, until falling off at high frequencies due to the moment-of-inertia (shaking the \gls{sc} requires more and more energies at high frequencies). The thruster noise of \gls{mps} widens the peak around $30\,\text{mHz}$ and \gls{dws} noise contributes to the fall-off behavior.

The lower panels show the \gls{mosa} jitter. The spectral shape is somewhat simpler: there is a maximum at low frequencies, with a power law fall-off towards higher frequencies. The \gls{mosa} $\eta$ jitter is very subdominant, even in its simplistic modeling. The $\phi$ jitter level is determined by the \gls{dws} readout noise level and the actuation force noise level. The actuation force noise level was set such that the peak does not exceed $2\,\unit[power-half-as-sqrt,per-mode=symbol]{\nano\radian\per\Hz\tothe{0.5}}$.

Additionally, reference lines were added describing jittes used in previous works. Here the comparison is done with \cite{hartig2025tilt} and \cite{george2023fisher}.  Both white jitters, with a low frequency roll-off, as well as coloured jitters are considered. The white jitter references are the same as in \cite{paczkowski_postprocessing_2022}, while the coloured jitters in \cite{hartig2025tilt} have been modified from \cite{george2023fisher}, case B, which in turn are originally from a dynamics simulation \cite{hewitson_lisasim_21}.

The white jitter reference \glspl{asd} used are identical for all \gls{sc} Cardan angles and given by
\begin{align}
	5\, \frac{\unit{\nano\radian}}{\sqrt{\unit{\Hz}}} \sqrt{1+\left(\frac{\SI{0.8}{\milli\Hz}}{f}\right)^4}\ , \label{eq:ref_white_sc_jitter}
\end{align}
while the white \gls{mosa} jitter \glspl{asd} have the same spectral shape, but level out to \SI[power-half-as-sqrt,per-mode=symbol]{1}{\nano\radian\per\Hz\tothe{0.5}} for $\eta$, and for $\phi$ to \SI[power-half-as-sqrt,per-mode=symbol]{2}{\nano\radian\per\Hz\tothe{0.5}}. 
\begin{figure}[!ht]
\centerline{\includegraphics[width=\linewidth, clip]{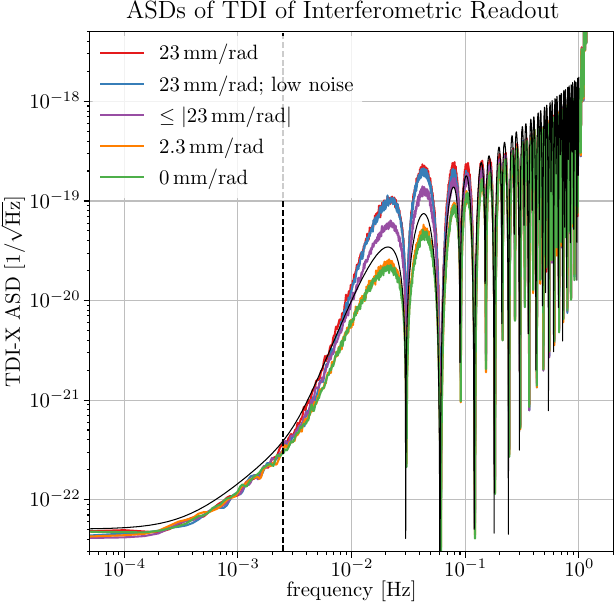}}
\caption{\glspl{asd} of \gls{tdi}-X (second-generation Michelson variables) for different simulation scenarios. These include different settings for the \gls{ttl} coefficients and different \gls{dws} noise levels (low readout noise means half of the nominal settings). 'Random' refers to the \gls{ttl} coefficients being sampled from a uniform distribution. The black line gives the noise requirement of \cite{paczkowski_postprocessing_2022} (c.f. Eqs.\myhyperref{eq:req-tm-displacement},\myhyperref{eq:req-tm-acc}). The vertical dashed line gives the drag-free frequency bandwidth \cite{inchauspe23_dynamics}.\\
The variation in \gls{dws} readout noise only has an impact on frequencies around \SI{100}{\milli\Hz}. For equal \gls{ttl} coefficients of \SI[per-mode=symbol]{2.3}{\milli\metre\per\radian} the shape of the \gls{asd} is very close to the simulation with no \gls{ttl} contribution. The \glspl{asd} are calculated from a \SI{e6}{\s} dataset with $N_\text{avg.}=50$ (c.f. App.~\myhyperref{app:lisanode}).}
\label{fig:tdi-output}
\end{figure}

The white jitter reference lines are significantly higher than the results of the closed-loop dynamics simulation. The white jitters can still be useful for requirement modeling. However, these do not provide realistic expectations of jitters. Note that the simulation here uses the same level of \gls{dws} readout noise of \SI[power-half-as-sqrt,per-mode=symbol]{0.2}{\nano\radian\per\Hz\tothe{0.5}} \cite{hartig2025tilt}.

The colored jitter \glspl{asd} are given for all \gls{sc} angles by
\begin{align}
	\begin{split}
	&0.7^2 \cdot 40\, \frac{\unit{\femto\radian}}{\sqrt{\unit{\Hz}}} \left(1 + \frac{f^2}{(\SI{20}{\micro\Hz})^2}\right) \left(1 + \frac{f^2}{(\SI{8}{\Hz})^2}\right)^{-1} \\
    &\quad \cdot\frac{(\SI{10}{\milli\Hz})^2}{\sqrt{0.7^2 \left(f^2 - (\SI{10}{\milli\Hz})^2\right)^2 + (\SI{10}{\milli\Hz})^2 f^2}} \\
	& \quad\cdot \frac{(\SI{50}{\milli\Hz})^2}{\sqrt{0.7^2 \left(f^2 - (\SI{50}{\milli\Hz})^2\right)^2 + (\SI{50}{\milli\Hz})^2 f^2}}  \ . 
	\end{split}\label{eq:ref_col_sc_jitter}
\end{align}
and for \gls{mosa} $\phi$ lowered by multiplying by \num{0.2}. The colored \gls{mosa} $\eta$ is set to zero in \cite{hartig2025tilt}, updating from \cite{george2023fisher}. The \gls{mosa} reference lines in Figure~3 of \cite{george2023fisher} can be reproduced by
\begin{align}
	B\, \sqrt{\frac{(\SI{10}{\nano\Hz})^2}{(\SI{10}{\nano\Hz})^2+ f^4}}\ , \label{eq:ref_col_mosa_jitter2}
\end{align}
using $B=\SI[power-half-as-sqrt,per-mode=symbol]{1}{\micro\radian\per\Hz\tothe{0.5}}$ for $\eta$ and $B=\SI[power-half-as-sqrt,per-mode=symbol]{10}{\micro\radian\per\Hz\tothe{0.5}}$ for $\phi$.

\begin{figure*}[!htb]
\centerline{\includegraphics[width=\linewidth, clip]{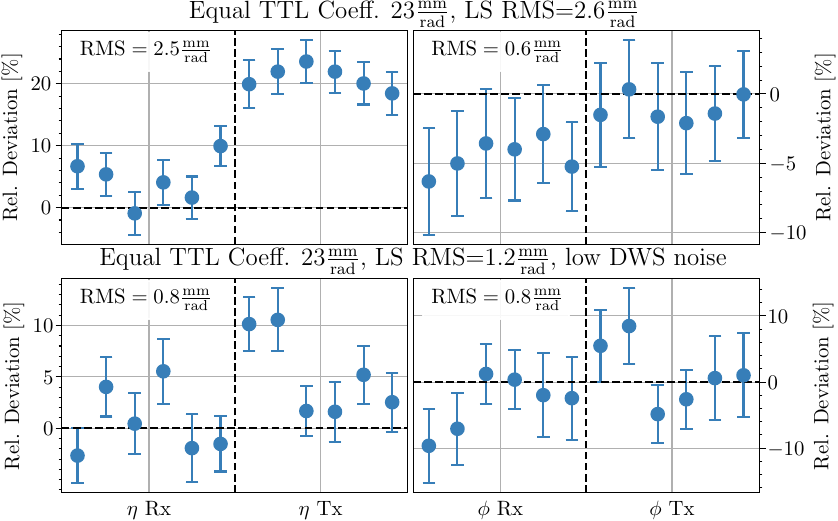}}
\caption{Inferred \num{24} \gls{ttl} parameters for two simulation, ordered in their categories as $\{12,13,23,21,31,32\}$. The top panels show results for the standard \gls{dws} noise settings of \SI[power-half-as-sqrt,per-mode=symbol]{0.2}{\nano\radian\per\Hz\tothe{0.5}} for the white noise in angles, the bottom panels with half the noise. The results are given in terms of a relative parameter deviation, i.e., the difference of the inferred parameter and the true parameter, normalized by the true parameter. The error bars are calculated from the standard deviations of \num{100} simulations (c.f. App.~\myhyperref{app:error}). The \gls{rms} errors reported are also split into contributions from the inferred $\eta$ and $\phi$ parameters. The results show that although the \gls{ttl} contribution is clearly visible for the larger \SI[per-mode=symbol]{23}{\milli\metre\per\radian} \gls{ttl} coefficients (c.f. Fig.~\myhyperref{fig:tdi-output}), the inference still has a substantial relative error of  $11.3\% $ for the standard readout noise, and $5.2 \%$ for the low \gls{dws} noise.}
\label{fig:ttl-inference-result}
\end{figure*}

This reference line in Eq.~\myhyperref{eq:ref_col_sc_jitter} is much closer to the closed-loop \gls{sc} jitters, but several details are different: the overall amplitude is ten times higher in the reference; the jitters for $\Theta_1$, $H_1$, $\Phi_1$ are not all the same; the control loop adds the \gls{dws} readout noise as a plateau for low frequencies; the fall-off towards higher frequencies is not a pure power-law.

For the \gls{mosa} jitters, the closed-loop simulated $\eta$ is lower in amplitude than the other references. For $\phi$, the white jitters overestimate the amplitude as well, while the dashed black line gives a different spectral shape to the simulated jitters. Lastly, the case B dash-dotted reference line from \cite{george2023fisher} has a similar high frequency behavior as the outputs of this work: there is a plateau at low frequencies, with power law fall off towards higher frequencies. But the reference line has an overall higher amplitude, and the plateau also switches to a power law at a lower frequency.

The next Fig.~\myhyperref{fig:jitters-sensing} shows the \gls{dws} outputs for one \gls{mosa}. Again, the left column shows the angle time series, the right the derivative. The top row shows the simulation output for the nominal \gls{dws} readout noise settings of $0.2\, \unit[power-half-as-sqrt,per-mode=symbol]{\nano\radian\per\Hz\tothe{0.5}}$, and the bottom row with half the amount. Note that the given number refers to the amplitude of the white noise in the angles.

Both the $\eta$ and $\phi$ output show the \gls{dws} readout noise as a plateau at high frequencies, where the control loop is not efficient anymore and thus cannot remove it. The $\phi$ output shows that it is essentially the sum of the \gls{sc} and \gls{mosa} $\phi$ jitters (c.f. Eqs.~\myhyperref{eq:dws-approx-phi1},\myhyperref{eq:dws-approx-phi2}). In the $\eta$ output, the \gls{mosa} $\eta$ jitter is so subdominant it cannot be seen directly. The outputs are a linear combination of the $\Theta$ and $H$ \gls{sc} jitters (c.f. Eqs.~\myhyperref{eq:dws-approx-eta1},\myhyperref{eq:dws-approx-eta2}). Comparing the standard and low \gls{dws} noise case, the peak around \SI{30}{\milli\Hz} is more pronounced with reduced noise. Generally, the $\phi$ output has a higher signal-to-noise ratio, as its underlying jitter levels are larger.

The last figure of this subsection, Fig.~\myhyperref{fig:tdi-output}, shows the full interferometric output after \gls{tdi} processing, for several different scenarios. Only the \gls{tdi}-X output (2nd generation Michelson observables) is shown, as Y and Z look very similar. The baseline scenario is a simulation without any \gls{ttl} coupling present. Beyond that, four coupling scenarios are shown: equal coefficients of \SI[per-mode=symbol]{23}{\milli\metre\per\radian} and \SI[per-mode=symbol]{2.3}{\milli\metre\per\radian}, random coefficients drawn from a uniform distribution $\mathcal{U}([-23,23]\,\unit[per-mode=symbol]{\milli\metre/\radian})$, and equal coefficients of \SI[per-mode=symbol]{23}{\milli\metre\per\radian} with reduced \gls{dws} readout noise.

The reference line is calculated with the requirements of \cite{paczkowski_postprocessing_2022} on the \gls{tm}-to-\gls{tm} displacement noise \gls{asd}
\begin{equation}
    13.5 \frac{\unit{\pico\m}}{\sqrt{\unit{\Hz}}} \sqrt{1+\left(\frac{\SI{2}{\milli\Hz}}{f}\right)^4} \label{eq:req-tm-displacement}
\end{equation}
and the residual acceleration noise \gls{asd} on a single \gls{tm}
\begin{equation}
    2.7 \frac{\unit{\femto\m}}{\unit{\s\tothe{2}}\sqrt{\unit{\Hz}}} \sqrt{1+\left(\frac{\SI{0.4}{\milli\Hz}}{f}\right)^2} \sqrt{1+\left(\frac{f}{\SI{8}{\milli\Hz}}\right)^4} \label{eq:req-tm-acc}
\end{equation}
using an analytic description of the \gls{tdi} variables, given in detail in \cite{babak2021lisasensitivitysnrcalculations}.

When calculating the \gls{asd} here, the \gls{psd} is divided by the central laser frequency to give the result in fractional frequency units. The figure shows that due to generally lower jitters, the case of \gls{ttl} coefficients of \SI[per-mode=symbol]{2.3}{\milli\metre\per\radian} is barely distinguishable from a simulation without any \gls{ttl} coupling. This is why in this publication, for the \num{1}-day datasets, often the case of \SI[per-mode=symbol]{23}{\milli\metre\per\radian} coefficients will be considered, such that a sufficient signal-to-noise ratio is guaranteed. Note that even when limiting the absolute value of the coefficients to $2.3\,\text{mm/rad}$, it is possible to have specific combinations that increase the \gls{ttl} contribution by roughly a factor \num[parse-numbers=false]{\sqrt{2}} (Fig.~6 of \cite{wanner2024depth}). The reduction of \gls{dws} readout noise does not have much of an effect, except for the changing the amplitude of the three peaks around \SI{100}{\milli\Hz}.

Lastly, note that the \gls{ttl} contribution to the full interferometric output after \gls{tdi} is confined to frequencies of \SI{8}{\milli\Hz} to $\sim$\SI{200}{\milli\Hz}. 

\subsection{Inference Results}\label{sec:results-inference}
\begin{figure*}[!ht]
\centerline{\includegraphics[width=\linewidth, clip]{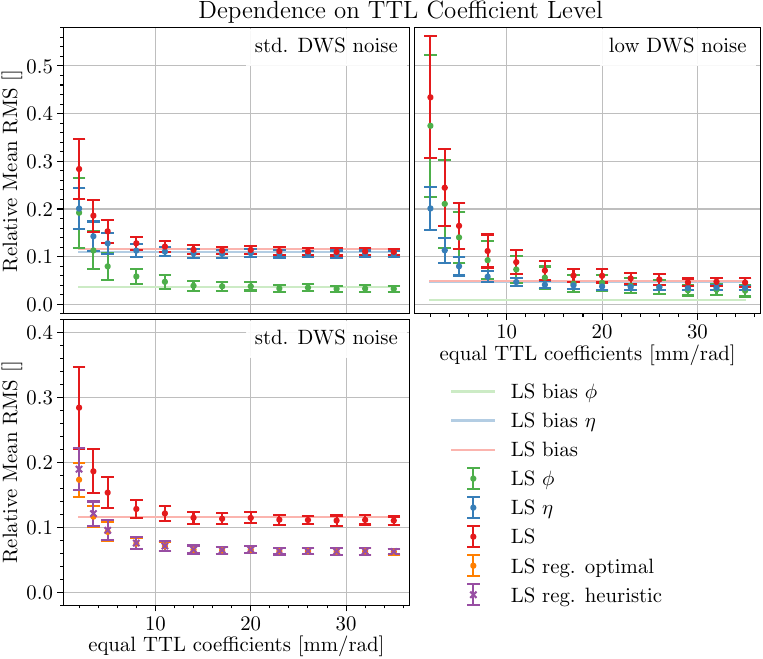}}
\caption{The panels show the change in the relative error when varying equal \gls{ttl} coefficient levels in the simulations. Each data point corresponds to the mean of the \gls{ttl} coefficient estimates' \gls{rms} error over \num{100} simulations, with the error bars given by the standard deviation. The upper left panel shows simulation results with nominal \gls{dws} noise settings of \SI[power-half-as-sqrt,per-mode=symbol]{0.2}{\nano\radian\per\Hz\tothe{0.5}}, the upper right panel with noise lowered to half. In the upper panels the results for the mean \gls{rms} values are also split into their contributions from the $\eta$ and $\phi$ parameters. Furthermore, the theoretical predictions for the bias of the \gls{ls} estimator are shown in lighter colors. The lower panel again shows the simulation with standard \gls{dws} noise levels, but now comparing the \gls{ls} results to the regularized \gls{ls} estimator, with the regularization parameter either chosen in the optimal way, or through the heuristic scheme described in Eq.~\myhyperref{eq:reg-heur-lam}.}
\label{fig:ttl-level-dependence}
\end{figure*}
Here we present the results of the inference pipeline for the basic scenario of using \num{1}-day datasets to infer the \gls{ttl} coupling coefficients without any additional excitation of \gls{mosa} or \gls{sc} \gls{dof}. We also do not consider any additional gravitational wave strains in the data, which has been handled by previous works \cite{hartig2025tilt,hartig2025postprocessing}. 

The \gls{dws} outputs and the full interferometric measurements are put through the \gls{tdi} algorithm as explained in the methods section. Then the output is filtered for frequencies where the jitter is dominant over the \gls{dws} readout noise (c.f. App.~\myhyperref{app:filtering}). This filtered data is then used to estimate the \gls{ttl} coefficients with the \gls{ls} estimator. 

The inference result for such a simulation and an additional simulation with reduced \gls{dws} readout noise can be seen in Fig.~\myhyperref{fig:ttl-inference-result}. The $y$-axis shows the relative deviation from the true \gls{ttl} parameters. The panels are split into $\eta$ and $\phi$, and Rx and Tx parameters. The order in each panel is $\{12,13,23,21,31,32\}$. Each row presents the results of a single simulation, where the error bars are calculated from the measured covariance matrix of \num{100} simulations.

In Fig.~\myhyperref{fig:ttl-inference-result} a broad theme is established: for the nominal \gls{dws} readout noise settings, the signal-to-noise ratio is worse in the $\eta$ \gls{dws} output compare to $\phi$, which is reflected in their individual \gls{rms} values. Moving to lower readout noise levels, the \gls{rms} value for the $\eta$ coefficients is improved, while the \gls{rms} values for the $\phi$ coefficients approximately stay at the same level. 

Especially in the case of the nominal readout noise settings, it is evident that the estimated error bars do not capture the full deviation from the true \gls{ttl} parameters. This is due to bias of the estimator (c.f. App.~\myhyperref{app:bias}), which is not taken into account in the estimation of the error bars.  

The bias is further investigated in Fig.~\myhyperref{fig:ttl-level-dependence}. It shows the relationship between the level of equal \gls{ttl} coefficients and the resulting \gls{rms} values. In the plots, every data point corresponds to the mean of \num{100} simulations, and the error bars are given by the standard deviation. The $y$-axis is reported as the relative mean \gls{rms}, i.e., the mean of the \gls{rms} values divided by the true \gls{ttl} coefficient level. The top left figure then shows that there is a linear relationship between the level of the equal \gls{ttl} coefficients and the \gls{rms} value, provided that the level is not so low that the signal-to-noise ratio in the full interferometric output gets deteriorated to a point where inference is not really possible anymore with the \gls{ls} algorithm. 

The plot then also shows the calculated bias of the estimator, which for equal \gls{ttl} coefficients only depends on the $K_{\alpha, 3}$ coefficients (c.f. App.~\myhyperref{app:bias}). The \gls{rms} value of the bias, for this scenario of equal coefficients $C_\text{eq.}$, is then
\begin{align}
    \begin{split}
    &\sqrt{\frac{1}{24}\left(\sum_{i=1}^{12} \left(4 K_{\eta 3} C_\text{eq.}\right)^2 + \sum_{i=1}^{12} \left(4 K_{\phi 3} C_\text{eq.}\right)^2\right)} \\
    =& \sqrt{8 (K_{\eta 3}^2+K_{\phi 3}^2)} \:|C_\text{eq.}|\ .
    \end{split}
\end{align}
In the relative mean \gls{rms} plot this is then a horizontal line, which very well explains the measured \gls{rms} values for sufficiently large \gls{ttl} coupling levels. The bias coming from the $\eta$ coefficients is much higher as the signal-to-noise ratio in the \gls{dws} output is lower. 

The right panel then shows the same variation of \gls{ttl} coefficient level, but for simulations with a reduced \gls{dws} readout noise level. It is evident there that for medium and larger \gls{ttl} coefficients, the \gls{rms} value is reduced as the signal-to-noise ratio in the $\eta$ channel is improved. As seen before in Fig.~\myhyperref{fig:ttl-inference-result}, the \gls{rms} value for the $\phi$ channel is largely unaffected. The increase of the \gls{rms} value towards lower \gls{ttl} coefficient values is stronger, which is due to higher levels of correlations (which is investigated in greater detail in the first experiment on \gls{dws} readout noise levels, Sec.~\myhyperref{sec:exp-1}).

The lower panel shows the results for the nominal noise settings, and compares this to the regularized \gls{ls} estimator. The parameter $\lambda$ needs to be chosen, and tested fixed-point iterations did not yield good results \cite{GOMESDEPINHOZANCO2025109820}. The optimal way of choosing $\lambda$ is by checking which one gives a minimal \gls{rms} value and then using this. This is however not possible in a real scenario, as the true \gls{ttl} parameters are not known. Thus we propose a heuristic way of determining it: for a given \gls{ttl} coupling and noise level, determine the mean optimal $\bar{\lambda}_\text{opt.}$. Extensive testing has shown that this $\bar{\lambda}_\text{opt.}$ is independent of the \gls{ttl} level, but not the noise level. Taking $\bar{\lambda}_\text{opt.}\approx \SI{1.64e-16}{\Hz\tothe{2}\radian\tothe{2}\per\mm\tothe{2}}$ from \num{100} simulations with the nominal \gls{dws} readout noise level, calculate the heuristic $\lambda_\text{heur.}$ by
\begin{align}
    \lambda_\text{heur.} = \min\!\left(\frac{n_\text{dws}^\text{ASD}}{0.2\,\unit[power-half-as-sqrt,per-mode=symbol]{\nano\radian\per\Hz\tothe{0.5}}}, \frac{1}{8}\right)\,\bar{\lambda}_\text{opt.}\ . \label{eq:reg-heur-lam}
\end{align}
This just linearly scales $\lambda_\text{heur.}$ with the relative noise level, down to a lower floor. In the lower panel of the figure it can be seen that both methods work similarly well, which shows how stable the choice of $\lambda$ is. The added constraint to the \gls{ls} estimator improves the \gls{rms} value by about a factor \num{1.5} in the case of equal \gls{ttl} coupling coefficients. The estimated values lie under the bias contribution, which is allowed as the computed bias is only valid for the unregularized estimator. 

\begin{figure}[!ht]
\centerline{\includegraphics[width=\linewidth, clip]{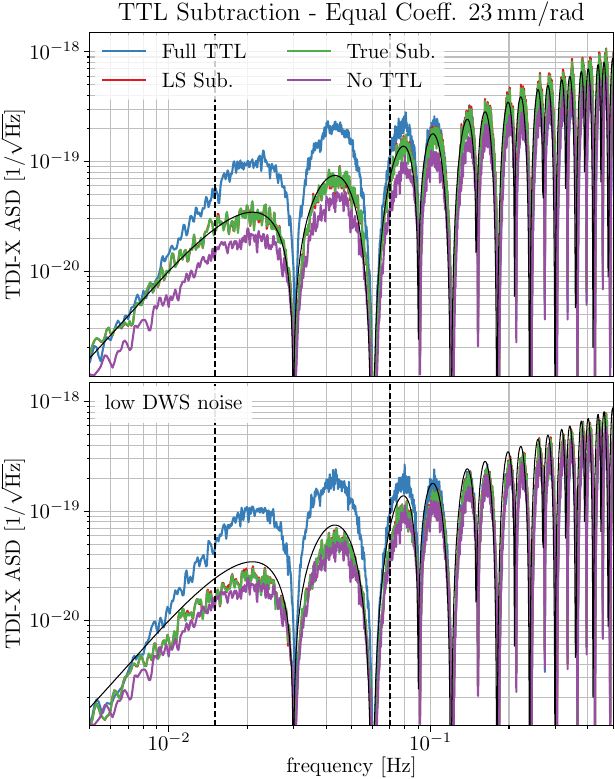}}
\caption{Demonstration of the suppression of the \gls{ttl} contribution to the noise. Both panels deal with the case of equal \gls{ttl} coefficients of \SI[per-mode=symbol]{23}{\milli\metre\per\radian}. The top panel has the standard \gls{dws} noise levels of \SI[power-half-as-sqrt,per-mode=symbol]{0.2}{\nano\radian\per\Hz\tothe{0.5}}, the bottom one with noise lowered to half. Both panels demonstrate that the \gls{ttl} coupling noise can be removed to the same level as when using the true coefficients. The black line gives the noise requirement of \cite{paczkowski_postprocessing_2022} (c.f. Eqs.\myhyperref{eq:req-tm-displacement},\myhyperref{eq:req-tm-acc}). However, in the case of nominal \gls{dws} readout noise in the top panel, additional noise is added in high frequency regime above \SI{200}{\milli\Hz}, increasing the total noise beyond the level before subtraction. This can be improved upon with further filtering. The \glspl{asd} are calculated with $N_\text{avg.}=10$ (c.f. App.~\myhyperref{app:lisanode}).}
\label{fig:ttl-sub-simple}
\end{figure}

As a final step, the estimated \gls{ttl} coefficients can be used to suppress the \gls{ttl} contribution to the total interferometric readout. The results can be seen in Fig.~\myhyperref{fig:ttl-sub-simple}.
The upper panel is for the nominal simulation, the lower one for one with reduced \gls{dws} noise. The interferometric output with full \gls{ttl} contribution is shown, with the subtractions of \gls{ttl} coupling using the \gls{ls}-inferred parameters and the true parameters. For reference, also a different simulation without any \gls{ttl} coupling is shown.

With nominal settings, the \gls{ttl} contribution can be removed up to the noise level in the \gls{dws} channels: the \gls{ttl} contribution in the interferometric readout comes without the \gls{dws} readout noise. There is barely any difference between the subtracted signal, no matter whether true or \gls{ls}-inferred parameters were used. Note that for high frequencies above \SI{200}{\milli\Hz}, subtracting the \gls{ttl} signal without any additional filters actually adds noise to the full interferometric output. As expected, for the reduced noise settings, the subtracted results are compatible with the simulation without any \gls{ttl} coupling.

\subsection{Experiment 1: DWS Noise Dependency}\label{sec:exp-1}
\begin{figure*}[!ht]
\centerline{\includegraphics[width=\linewidth, clip]{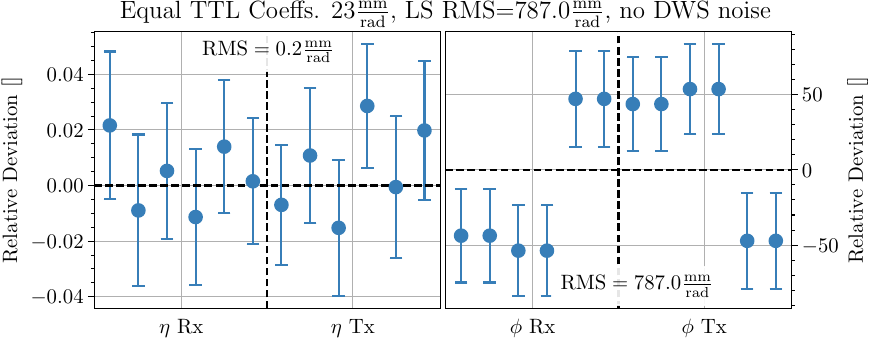}}
\caption{Inferred \num{24} \gls{ttl} parameters for a single simulation with zero \gls{dws} readout noise, ordered in their categories as $\{12, 13, 23, 21, 31, 32\}$. The results are given in terms of a relative parameter deviation, i.e., the difference of the inferred parameter and the true parameter, normalized by the true parameter. The error bars are calculated from the standard deviations of \num{100} simulations (c.f. App.~\myhyperref{app:error}). The \gls{rms} errors reported are also split into contributions from the inferred $\eta$ and $\phi$ parameters. \\
Compared with non-zero \gls{dws} noise settings in Fig.~\myhyperref{fig:ttl-inference-result}, the inference for the $\eta$ parameters has improved due to a higher signal-to-noise ratio in the data. However, due to high correlations for the $\phi$ parameters, the estimator can constrain only certain combinations of parameters (c.f. App.~\myhyperref{app:corr}). With only subdominant \gls{mosa} jitter and no noise, the \gls{dws} $\phi$ outputs become virtually the same for the left and right \gls{mosa} of a single \gls{sc}. This leads to an ill-conditioned estimation.}
\label{fig:ttl-inference-result-no-dws}
\end{figure*}
\begin{figure*}[!ht]
\centerline{\includegraphics[width=\linewidth, clip]{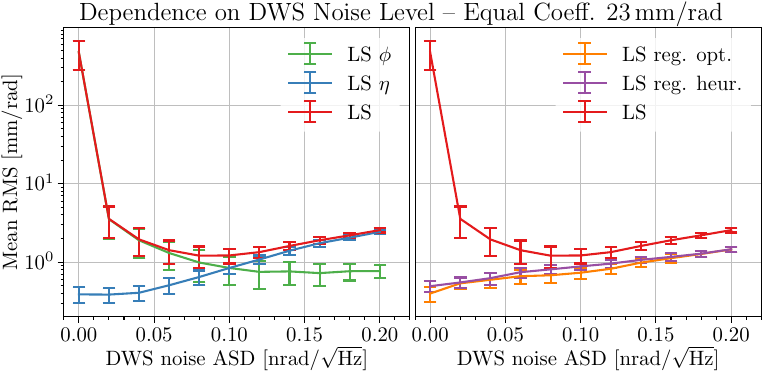}}
\caption{The panels show the change in the \gls{rms} error when varying \gls{dws} noise levels in the simulations. Each data point corresponds to the mean of the \gls{ttl} coefficient estimates' \gls{rms} error over \num{100} simulations, with the error bars given by the standard deviation. All simulations were run with equal \gls{ttl} coefficients of \SI[per-mode=symbol]{23}{\milli\metre\per\radian}. The left panel shows the results of the \gls{ls} estimator, also divided into the contributions from the $\eta$ and $\phi$ inference. The right panel compares the \gls{ls} results to the regularized \gls{ls} estimator, with the regularization parameter either chosen in the optimal way, or through the heuristic scheme described in Eq.~\myhyperref{eq:reg-heur-lam}. The left figure show the expected decrease of the inference in the $\eta$ coefficients with less \gls{dws} noise. The $\phi$ coefficient estimates increase dramatically due to increased correlations between the left and right \gls{mosa} in each \gls{sc}. The right panel shows that this can be prevented using a regularization term in the \gls{ls} estimator.}
\label{fig:ttl-dws-dependence}
\end{figure*}
In this first experiment with the simulation, we want to investigate the dependence on the \gls{dws} readout noise levels. First of all, the case with no readout noise whatsoever is interesting to look at. Figure~\myhyperref{fig:ttl-inference-result-no-dws} shows the results of the inference. Compared to the regular simulation in previous Fig.~\myhyperref{fig:ttl-inference-result}, the estimation of the $\eta$ \gls{ttl} parameters works far better. However, the $\phi$ parameters can only be estimated with uncertainties many times larger than the coefficient value, but there is a pattern to them: we seemingly have $C_{ij\phi}^\text{Rx}=C_{ik\phi}^\text{Rx}$, and $C_{ij\phi}^\text{Rx}=-C_{ij\phi}^\text{Tx}$ for $(i,j,k)=(1,2,3)$. This is due to a perfect correlation of the \gls{dws} $\phi$ outputs on one \gls{sc}, which leads to this correlation pattern after \gls{tdi}. This is further investigated in App.~\myhyperref{app:corr}.

This pattern is then also found in Fig.~\myhyperref{fig:ttl-dws-dependence}. On the $x$-axis, the \gls{dws} readout noise levels are varied between zero and the nominal scenario of \SI[power-half-as-sqrt,per-mode=symbol]{0.2}{\nano\radian\per\Hz\tothe{0.5}}. Every data point corresponds to the mean of \num{100} simulations, and the error bars to the standard deviation. In the left panel, the $\eta$ parameters show the expected behavior: for lower noise, the inference has a lower \gls{rms} error as well. The $\phi$ parameters do not exhibit this behavior, as for lower noise levels, the \gls{dws} channels within a \gls{sc} become more tightly correlated, leading to higher correlations in the \gls{ttl} parameters (c.f. App.~\myhyperref{app:corr}). 

The right panel then shows the results for the regularized \gls{ls} estimator, with the single parameter $\lambda$ chosen according to Eq.~\myhyperref{eq:reg-heur-lam} in the heuristic case. This then lowers the \gls{rms} error by a factor of about \num{1.5} in the cases where the noise is still large enough to avoid major correlations. When the readout noise approaches zero and the \gls{rms} error of the standard \gls{ls} estimator starts to diverge, the regularization scheme works well and manages to stay close to the true parameters. 
\begin{figure*}[!ht]
\centerline{\includegraphics[width=\linewidth, clip]{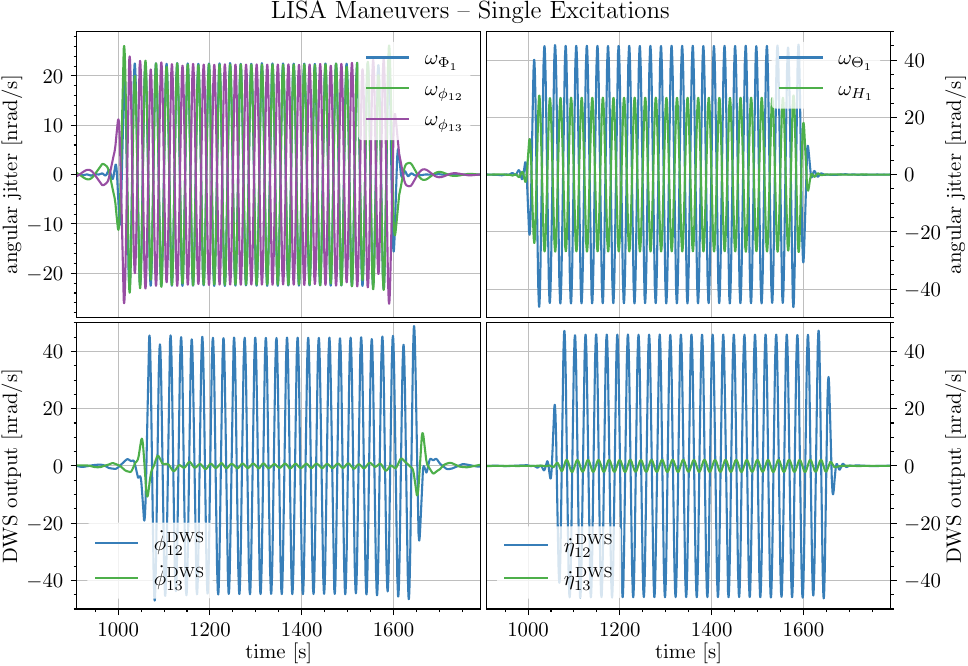}}
\caption{The panels show how $\eta$ (left) and $\phi$ (right) maneuvers are induced. The top panels display the actual rotations induced in the \gls{sc} and \gls{mosa} \gls{dof}. The bottom panels show the resulting \gls{dws} readout, with excitations only present on the left \gls{mosa}. Details on the required excitations can be found in Sec.~\myhyperref{sec:exp-2-man}. Maneuver duration is \SI{600}{\s}. The shift between top and bottom panels in the start of the excitations is due to the delay caused by the \gls{adc}. A bandpass filter has been applied to the displayed data with frequencies \SI{15}{\milli\Hz} and \SI{70}{\milli\Hz}.}
\label{fig:ttl-man-single-sc-combined}
\end{figure*}

As a last point in this experiment, note that while the \gls{rms} error of the inferred parameters is large compared to the true parameters, subtraction of the \gls{ttl} signal is still possible. This is because the strong correlations in the estimated coefficients lead to cancellations in the linear combination $\dot{\myvec{\alpha}}_\text{TDI}^\text{dws}\cdot \hat{\myvec{C}}_\text{TTL}$. To quantify this, compare the time series of the interferometric output, with the \gls{ttl} signal subtracted by the \gls{ls}-inferred and true parameters, after applying a lowpass filter with \SI{70}{\milli\Hz} to remove high-frequency noise. Over the \num{1}-day dataset, the \gls{rms} deviation is \SI{0.064}{\micro\Hz}, while the maximum deviation is \SI{0.27}{\micro\Hz}. The \gls{rms} value of the subtraction with the true parameters is \SI{1.7}{\micro\Hz}. Hence, depending on whether the \gls{ttl} coefficients themselves need to be known (e.g. to minimize \gls{ttl} coupling using the \gls{bam}), or just the \gls{ttl} signal needs to be removed from the data stream, it is necessary to add a regularization term to the \gls{ls} estimator.

\subsection{Experiment 2: Maneuvers} \label{sec:exp-2-man}
\begin{figure*}[!ht]
\centerline{\includegraphics[width=\linewidth, clip]{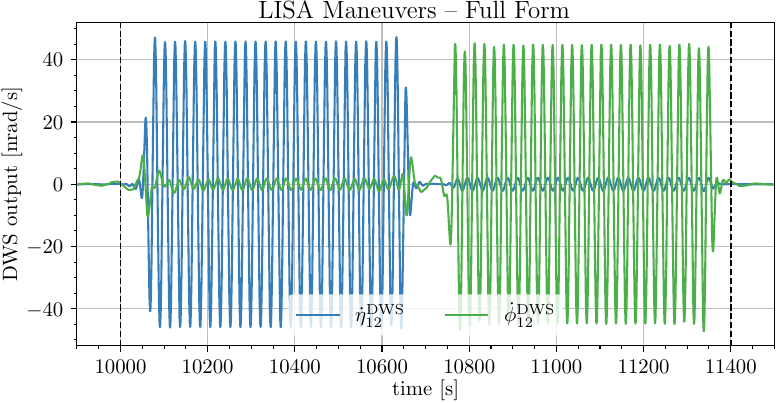}}
\caption{The \gls{dws} outputs for \gls{mosa} 12 during a full maneuver as defined in Sec.~\myhyperref{sec:exp-2-man}. The maneuver consists of two single \SI{600}{\s} phases after each other, with a waiting time of \SI{100}{\s}. For the first part, the left \gls{dws} channel is excited in $\eta$ and the right in $\phi$, and in the second phase vice-versa. The frequencies of the excitations have been carefully chosen according to \cite{wegener2025design}. A bandpass filter has been applied to the displayed data with frequencies \SI{15}{\milli\Hz} and \SI{70}{\milli\Hz}. The vertical dashed lines correspond to the time interval chosen for parameter inference.}
\label{fig:ttl-full-man-dws-out}
\end{figure*}
As a second experiment, we deliberately introduce rotations in the \gls{sc} and \glspl{mosa} to excite the \gls{ttl} coupling. Several proposals exist for these so-called maneuvers, here we will implement the ones proposed in \cite{wegener2025design}.

As a first step, it is important to be able to excite single \gls{dws} output channels. The $\eta$ maneuvers are relatively easy to implement. As the \gls{sc} $\Theta$ jitter enters with opposite signs for adjacent \glspl{mosa} (c.f. Eq.~\myhyperref{eq:dws-approx-eta1-equal}), it is possible to excite one channel without the other. For example, to excite $\eta_{12}^\text{dws}$, excite $H$ with a sine curve with amplitude $A$, and then $\Theta$ with an amplitude $-\sqrt{3}A$. Then, neglecting noise and \gls{mosa} jitter, we have the excitations
\begin{align}
    &\eta_{12}^\text{dws} \approx \sqrt{3} A \sin(2\pi f_\text{guide} t)\ , \\
    & \eta_{13}^\text{dws} \approx 0\ .
\end{align}
There is of course a correction due to the breathing angle, but the correction is small and does not affect the result much. As the dynamics is implemented here as a closed-loop system, certain errors are present which cannot be seen in the original work's open-loop implementation \cite{wegener2025design}.

The $\phi$ maneuvers are a bit more tricky as the \glspl{mosa} in this simulation can only be actuated in a correlated fashion: $t_{\mathcal{H}_\myindex{1}}^\text{act.} = -t_{\mathcal{H}_\myindex{2}}^\text{act.}$ for the torques. This is because the measured \gls{dws} output is $4$-dimensional per \gls{sc}, but the underlying state space is $5$-dimensional ($\Theta$, $H$, $\Phi$, $\phi_1$, $\phi_2$). Thus here only $\phi_1 - \phi_2$ is controlled with the \gls{mosa} actuation. 

It is still possible to implement a maneuver: excite the \gls{mosa} \gls{dof} and then use the \gls{sc} $\Phi$ motion to compensate (c.f. Eq.~\myhyperref{eq:dws-approx-phi1}). Depending on the sign of the $\Phi$ excitations, either the left or right \gls{dws} $\phi$ output is excited, while the other is zero. 

As mentioned before, here we work with frequencies, not phases. But all maneuvers carry directly over to the derivatives. Additionally, in \cite{wegener2025design} a maximum amplitude is derived based on the \gls{sc} dimensions, weight, and maximum thruster force output. As the \gls{sc} is modeled differently here, this leads to a different $\SI{0.4}{\nano\radian}\cdot (1\,\unit{\Hz}/f)^2$. In terms of the derivative, this corresponds to $A_\text{max}\approx \SI{2}{\nano\radian} \cdot(\SI{1}{\Hz\tothe{2}}/f)$. For a frequency of \SI{40}{\milli\Hz} this gives \SI[per-mode=symbol]{50}{\nano\radian\per\second}.

In Fig.~\myhyperref{fig:ttl-man-single-sc-combined} the implemented maneuvers are shown. The excitations are implemented through an injected guidance signal of the controller. The injection is such that the nominal amplitude in $H$ is \SI{100}{\nano\radian}, and appropriately scaled for all other channels. These excitations are in the upper part of the possible thruster performance as described in the last paragraph. The left panels shows the $\phi$ maneuver with a frequency of \SI[parse-numbers=false]{43.\bar{3}}{\milli\Hz}, and the right the $\eta$ maneuver with the same frequency. The top row shows the internal states during the maneuver, the bottom row the \gls{dws} outputs. Note that the shifted start of the maneuver between top and bottom row is due to the \acrfull{adc}'s induced delay.

The data in the figure was passed through a bandpass filter to remove the high-frequency \gls{dws} readout noise and a low-frequency drift in the \gls{mosa} $\phi$ angles when actuating them, such that the sinusoidal signal and its residual in the adjacent \gls{mosa} can be seen better. The \gls{mosa} controller reacted with an initial excitation of about \SI[per-mode=symbol]{25}{\nano\radian\per\s} to the sudden start of the guidance signal, which then slowly drifts towards zero over the duration of the excitation. As the data for \gls{ttl} coupling inference is also filtered, this does not change the results there.

The maneuver is excited exactly for \SI{600}{\s}. It can be seen in the figure that the maneuver needs time to equilibrate to the intended sine excitation. Furthermore, while the excitation on the neighboring \gls{dws} is strongly suppressed, it is not zero. This residual excitation is phase-shifted with respect to the sine, and is due to the controller/actuator also having a phase-delayed component in their response.

The full maneuver then consists of two phases: excitations on all \gls{sc} for \SI{600}{\s}, then a waiting period of \SI{100}{\s} such that the excitation is also over after \gls{tdi}, then a second round of excitations for \SI{600}{\s}. The excitations are chosen in such a way that the resulting signals are uncorrelated after \gls{tdi}. This is achieved by choosing different frequencies for the sine waves, and by only exciting \lq uncorrelated\rq\ pairs. The resulting design uses three frequencies $f_1=43.\bar{3}\,\unit{\milli\Hz}$, $f_2=41.\bar{6}\,\unit{\milli\Hz}$, and $f_3=44.7\,\unit{\milli\Hz}$ \cite{wegener2025design} and is defined by:
\begin{itemize}
    \item[]Phase \num{1}: $\eta_{12}$,$\eta_{32}$ at $f_1$; $\eta_{21}$,$\phi_{31}$ at $f_2$; $\phi_{13}$,$\phi_{23}$ at $f_3$
    \item[]Phase \num{2}: $\phi_{12}$,$\phi_{32}$ at $f_3$; $\phi_{21}$,$\eta_{31}$ at $f_2$; $\eta_{13}$,$\eta_{23}$ at $f_1$
\end{itemize}
This is shown in Fig.~\myhyperref{fig:ttl-full-man-dws-out} for the \gls{mosa} 12 \gls{dws} output. The vertical dashed lines shows the time range out of the \num{1}-day dataset used for inference. The actually used range is larger, as the filtering introduces ringing that needs to be cut away (\SI{750}{\s} each end).
\begin{figure*}[!ht]
\centerline{\includegraphics[width=\linewidth, clip]{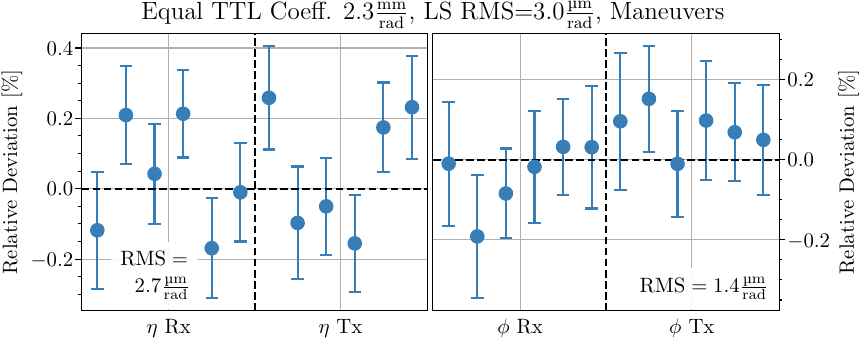}}
\caption{Inferred \num{24} \gls{ttl} parameters for a single simulation with a full maneuver, ordered in their categories as
$\{12, 13, 23, 21, 31, 32\}$. The results are given in terms of a relative parameter deviation, i.e., the difference of the inferred parameter and the true parameter, normalized by the true parameter. The error bars are calculated from the standard deviations of \num{100} simulations (c.f. App.~\myhyperref{app:error}). The \gls{rms} errors reported are also split into contributions from the inferred $\eta$ and $\phi$ parameters. Compared with simulations without maneuvers in Fig.~\myhyperref{fig:ttl-inference-result}, the inference works far better with the \gls{ls} estimator due to the high signal-to-noise ratios in both the interferometer and \gls{dws} outputs during the maneuver time. This works even though the equal \gls{ttl} coefficients have been reduced to the nominal value of \SI[per-mode=symbol]{2.3}{\milli\metre\per\radian}.}
\label{fig:ttl-inference-result-man}
\end{figure*}\begin{figure*}[!ht]
\centerline{\includegraphics[width=\linewidth, clip]{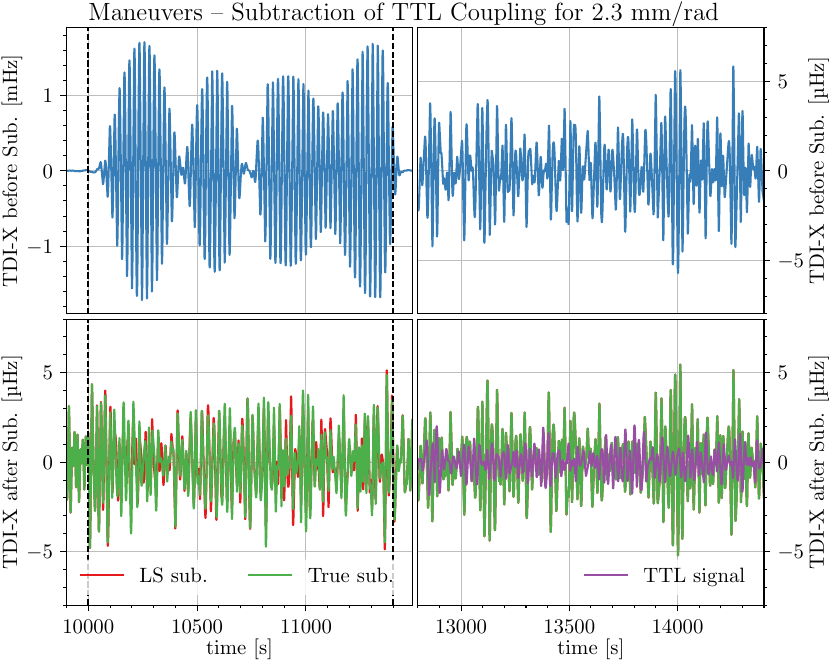}}
\caption{Demonstration of the removal of the \gls{ttl} contribution in the case of a full maneuver. The data is shown in the time domain (with a bandpass filter applied, using frequencies \SI{15}{\milli\Hz} and \SI{70}{\milli\Hz}), with the left panels showing the time during the maneuver, and the right panels some later time without any injected signal. The top panels show the interferometric output after \gls{tdi} but before subtraction, the bottom panels show the result after subtraction of the \gls{ttl} contribution. Outside the injection time, the subtraction with the \gls{ls}-estimated coefficients is nearly identical to the case when using the true coefficients. During the maneuver, even these small deviations get amplified by the large signal and thus the subtractions differ (during injection: \gls{rms} deviation of \SI{0.28}{\micro\Hz}; maximum deviation of \SI{1.1}{\micro\Hz}). For comparison, the lower right panel also shows the amplitude of the \gls{ttl} signal that is subtracted. For this simulation with equal \gls{ttl} coefficients of \SI[per-mode=symbol]{2.3}{\milli\metre\per\radian}, the \gls{ttl} signal is relevant, but not dominant.}
\label{fig:ttl-full-man-subtract}
\end{figure*}
\begin{figure*}[!ht]
\centerline{\includegraphics[width=\linewidth, clip]{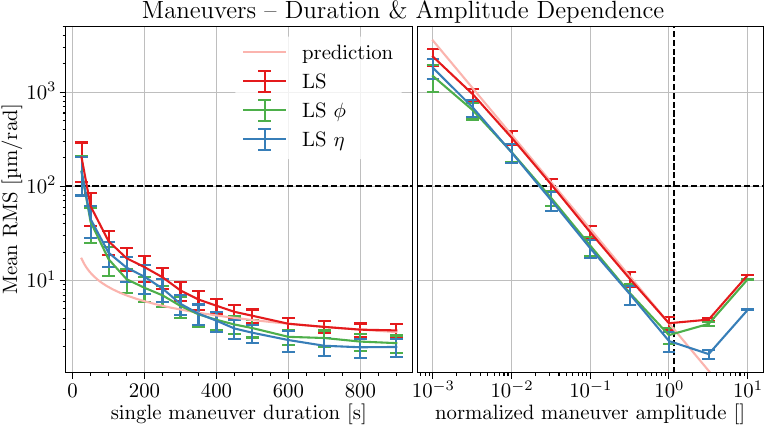}}
\caption{The panels show the change in the \gls{rms} error when varying duration and amplitude of a full maneuver. Each data point corresponds to the mean of the \gls{ttl} coefficient estimates' \gls{rms} error over \num{100} simulations, with the error bars given by the standard deviation. All simulations were run with equal \gls{ttl} coefficients of \SI[per-mode=symbol]{2.3}{\milli\metre\per\radian}. The left panel shows the results when changing the maneuver duration, the right panel the reaction to a changing amplitude. In both cases the results are displayed also in terms of the separate $\eta$ and $\phi$ contribution to the total \gls{rms} value. Also, a theoretical scaling curve is displayed in the lighter color. The amplitude of this prediction is fixed at the nominal cases of a duration of \SI{600}{\s} and the standard amplitude. The horizontal dashed line gives the usual target accuracy of the coupling coefficients at \SI[per-mode=symbol]{0.1}{\milli\metre\per\radian}.\\
The scaling law describes the reaction to a changing amplitude maneuver well, but underestimates the error for smaller maneuver duration, where the additional effect of a smaller effective amplitude increases the error. In the right panel, the normalized maneuver amplitude is with respect to the nominal injection amplitude, chosen such that a $H$ excitation amplitude of \SI{100}{\nano\radian} is achieved. The vertical dashed line gives the maximum possible maneuver amplitude given the expected thruster power. The rise towards higher amplitudes can be attributed to a cross-coupling with the \gls{tm} \gls{dof}.}
\label{fig:ttl-full-man-dur-amp-dependence}
\end{figure*}

The regular pipeline for inference is then applied to this \SI{1400}{\s} dataset. It was tested how sensitive the results are with respect to the choice of start and end points. The result is very stable, thus this is not optimized further. The inference result is shown in Fig.~\myhyperref{fig:ttl-inference-result-man}. The simulation was performed with equal \gls{ttl} coefficients of \SI[per-mode=symbol]{2.3}{\milli\metre\per\radian}. The \gls{rms} error of the inferred parameters is drastically lower as for the \num{1}-day datasets without maneuvers (c.f. Fig.~\myhyperref{fig:ttl-inference-result}). The bias of the estimator plays no role anymore, as the signal-to-noise ratio in the \gls{dws} output is very high, and correlations pose no problem as the maneuvers have been designed in a way to eliminate them, which holds up in this closed-loop dynamics setting. Further information on the correlations can be found in App.~\myhyperref{app:corr}.

The inferred \gls{ttl} coefficients can then be used again to remove the \gls{ttl} signal from the main interferometer output. The procedure is as before, only that now the parameters from the short \SI{1400}{\s} inference are used for the whole $1$-day dataset. The result can be seen in Fig.~\myhyperref{fig:ttl-full-man-subtract}. The upper row shows the original interferometer output after \gls{tdi} for two time windows (with maneuver and without), while the lower row shows the signal after \gls{ttl} subtraction. The dashed vertical lines again show the time window used for parameter inference. 

The signal during the maneuver in the upper left plot is large, measured against the interferometer output (\unit{\milli\Hz} vs. \unit{\micro\Hz}). It is however still possible to remove this imprint of the maneuver from the output, as can be seen in the lower left plot. A difference in the subtraction can be seen between the true and inferred \gls{ttl} parameters, even though they agree very well. This is just because the original maneuver signal is so large. So the science output of the interferometer could still be used during such a maneuver, except for very small \gls{gw} signals.

In the time window without maneuvers in the upper right plot, there is no large contribution from \gls{ttl} coupling as the equal coefficients just have a level of \SI[per-mode=symbol]{2.3}{\milli\metre\per\radian}. The \gls{ttl} signal can however still be removed, which can be seen in the lower right plot. As the true and inferred parameters agree so well, the subtraction is identical for both.

Lastly, it is interesting to investigate the dependence of the inferred parameters on the duration of a single phase of the maneuver, and the maneuver amplitude. This is shown in Fig.~\myhyperref{fig:ttl-full-man-dur-amp-dependence}. Every point corresponds to the mean of \num{100} simulations, and the error bars represent the standard deviation. These behaviors were also investigated in the original publication \cite{wegener2025design}. Looking at the estimated correlation matrix, it was possible to extract the relations $\operatorname{RMS}\sim A_\text{man.}^{-1}$ and $\operatorname{RMS}\sim T_\text{man.}^{-1/2}$ for the maneuver amplitude $A_\text{man.}$ and the duration of one phase of the maneuver $T_\text{man.}$ \cite{wegener2025design}. 

The dependence on the maneuver duration roughly follows this theoretical line (amplitude of $T^{-1/2}_\text{man.}$ determined at \SI{600}{\s}). Especially for low durations this underestimates the true \gls{rms} error by about a factor of \num{10}. This is because the closed-loop system needs time to attain the full amplitude, and for smaller durations this is not enough any more. So for shorter durations there is compound effect of not only reducing the duration, but also the effective amplitude. However, even for short maneuvers, the \gls{rms} error of the parameters is still below the target accuracy of \SI[per-mode=symbol]{0.1}{mm/rad}. 

The amplitude dependence follows the predicted curve very closely (amplitude of curve determined at nominal maneuver amplitude). Here it is clear that the maneuver amplitude can only be lowered by a value of about \num{20} before the target accuracy line is crossed. Interestingly, for larger amplitudes the \gls{rms} errors start increasing again. This is not because the maximum possible amplitude is surpassed, as the current implementation does not know about this limitation. Rather this is a cross-coupling of the movement of the \gls{sc} and \glspl{mosa} into the longitudinal \gls{dof} of the \glspl{tm}. This is mitigated by the coupling terms $\trafo{H}{B}\sum \myvec{f}_B^\mathcal{B}/m_B$ and $\trafo{H}{B} \myvec{r}_\myindex{H/B}^\myindex{\mathcal{B}} \times \trafo{H}{B}\vecwithrefrel{\dot{\omega}}{B}{O}{B}$ (c.f. Eq.~(69) in \cite{inchauspe23_dynamics}), with $\sum \myvec{f}_B^\mathcal{B}$ the forces applied to the \gls{sc}, $m_B$ it mass, and $\myvec{r}_\myindex{H/B}^\myindex{\mathcal{B}}$ the vector connecting the \gls{sc} center of mass $B$ with the \gls{tm} housing center $H$. Other terms only contribute at a four order-of-magnitude lower level. This relative motion of the \glspl{tm} is then picked up by the test-mass interferometer. For such large excitations, \gls{tdi} cannot fully remove the dynamics signal \cite{inchauspe23_dynamics}. The small residual is then an effective coupling term that is not accounted for in the \gls{ttl} model here. Thus the inferred \gls{ttl} parameters start to deviate from the true ones in order to try to maximally remove the \gls{ttl} signals. Also note that for these high maneuver amplitudes, non-linear terms in the \gls{dws} modeling start to become relevant, such that the purely linear \gls{ttl} coupling model would need to be adapted for this as well.
\begin{figure*}[!ht]
\centerline{\includegraphics[width=\linewidth, clip]{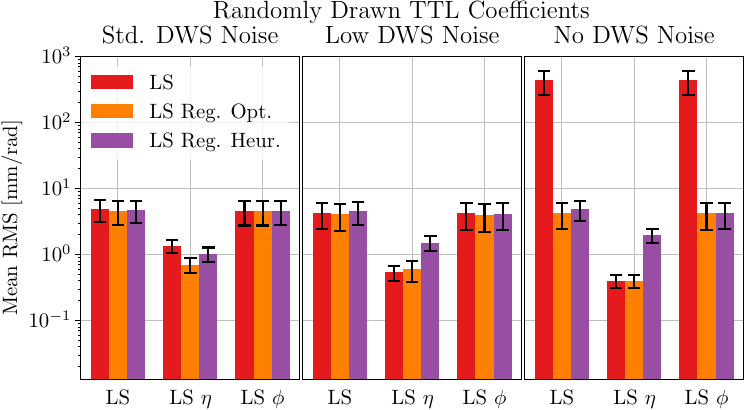}}
\caption{Results of inference when running simulations with randomly sampled \gls{ttl} coefficients from a uniform distribution on the interval $[-23,23]\,\unit[per-mode=symbol]{\milli\metre\per\radian}$. Each bar corresponds to the mean of the \gls{ttl} coefficient estimates' \gls{rms} error over \num{100} simulations, with the error bars given by the standard deviation. The result is split up in terms of the contributions of $\eta$ and $\phi$ to the total \gls{rms} error, and the different estimators: regular \gls{ls}, and regularized \gls{ls} with the optimally and heuristically chosen regularization parameter $\lambda$. The three panels show the results with respect to different \gls{dws} readout noise levels in the simulations: nominal (\SI[power-half-as-sqrt,per-mode=symbol]{0.2}{\nano\radian\per\Hz\tothe{0.5}}), lowered to half, and no readout noise. The panels show that for random \gls{ttl} coefficients, the inference task becomes somewhat harder. The coefficient estimate is worse than for the corresponding simulation with equal coefficients (c.f. Fig.~\myhyperref{fig:ttl-level-dependence}), and the regularization does not improve the result anymore. However, regularized \gls{ls} still is better in the highly-correlated, no \gls{dws} noise case, improving the inference result drastically.}
\label{fig:ttl-rand-coeff}
\end{figure*}

Appendix~\myhyperref{app:maneuver} investigates a different set of maneuvers mentioned in \cite{wegener2025design}. Additionally, the inference results for randomly drawn \gls{ttl} coefficients are shown as well as the impact of the breathing angles of the constellation instead of just using $\pi/6$.

\subsection{Experiment 3: Random TTL Coefficients}\label{sec:exp-3}
As a last experiment, we want to investigate the more realistic case of not just choosing equal \gls{ttl} coefficients, but instead drawing them from a uniform distribution. Here we will focus on the \num{1}-day datasets without maneuvers.

Figure~\myhyperref{fig:ttl-rand-coeff} shows the results of \num{100} simulations with randomly drawn \gls{ttl} coefficients in three different scenarios: nominal \gls{dws} readout noise levels, reduced levels to half, and no noise. The uniform distribution for the parameters is $\mathcal{U}([-23,23]\,\unit[per-mode=symbol]{mm/rad})$. 

It can be seen that the inference with the random coefficients seems to be harder. This is due to added correlations in the $\phi$ parameters introduced by the bias (c.f. App.~\myhyperref{app:corr}), as now the \gls{rms} error is dominated by the $\phi$ contribution. This pattern also stays the same when lowering the readout noise. Adding a regularization term to the \gls{ls} estimator also does not help anymore in improving the inference result. Only in the case of no readout noise is the regularization scheme useful in reducing the \gls{rms} error.

In App.~\myhyperref{app:maneuver} the random \gls{ttl} coefficients together with maneuvers are investigated. As the maneuvers are implemented in such a way to reduce correlations, the move from equal to random coefficients does not affect the inference result.

\section{Conclusion and Outlook}
This work provides detailed angular jitters of \gls{mosa} and \gls{sc} with modeled \gls{dws} output, all from a closed-loop dynamics. The simulation includes non-linear, and time-varying terms to be able to model \gls{mosa} jitter and the breathing of the constellation. These are then used to describe \gls{ttl} coupling within LISANode \cite{bayle_lisanode}. On the basis of a linear \gls{ttl} model adapted for working with frequency changes, this data is used to infer the \gls{ttl} coupling parameters. For this estimation, the \gls{ls} estimator with a possible regularization term is used on the time series data. 

Based on the usual estimate of all coupling coefficients being equal to \SI[per-mode=symbol]{2.3}{\milli\metre\per\radian}, the overall \gls{ttl} contribution to the full interferometric output after \gls{tdi} is not so dominant as previously thought. Specifically, it is limited to a frequency band of \SI{8}{\milli\Hz}--\SI{200}{\milli\Hz}, and the amplitude reaches the level of the other instrumental noises for the case of equal coupling coefficients (c.f. Fig.~\myhyperref{fig:tdi-output}). 

For the inference task, two scenarios were considered: a one-day dataset with just the instrumental noise, and a one-day dataset with $\sim$\SI{20}{\min} maneuvers creating sinusoidal angular excitations, based on the publication \cite{wegener2025design}.

For the first type of dataset, inference with such small \gls{ttl} contributions does not work very well. Thus, larger coefficients of \SI[per-mode=symbol]{23}{\milli\metre\per\radian} are used for this part. It is then possible to estimate the coefficients roughly to a $10\%$ level, based on the \gls{rms} error. The inference is impeded by a large bias contribution to the estimator, which has been extended to the case considered here, based on the publication \cite{hartig2025tilt}. Counter-intuitively, for lower \gls{dws} readout noises the \gls{rms} starts to diverge. This is due to correlations between the $\phi$ channels within one spacecraft. 

The dependence on \gls{ttl} coefficient level and noise levels have been thoroughly simulated (c.f. Figs.~\myhyperref{fig:ttl-level-dependence},\myhyperref{fig:ttl-dws-dependence}). Moving to coupling coefficients randomly drawn from a uniform distribution makes the inference task slightly harder, as the bias introduces additional correlations between estimated coefficients.

The second type of dataset, including specific maneuvers of the angular \gls{dof}, makes the inference task easy. With a strong signal-to-noise ratio in both the \gls{dws} output and the \gls{ttl} signal, the \gls{rms} error is on the level of $0.1\%$. The design of the maneuvers prevents correlations between \gls{ttl} parameters, which also holds up in this more realistic, closed-loop setting. Duration and excitation amplitude dependence are checked to be roughly in line with theoretical predictions from \cite{wegener2025design} (c.f. Fig.~\myhyperref{fig:ttl-full-man-dur-amp-dependence}) and can be used to find the optimal maneuver design for a given accuracy requirement. For large maneuver amplitudes, dynamical cross-couplings to the \gls{tm} \gls{dof} are excited. As this is not accounted for in the usual \gls{ttl} coupling model, the inferred parameters get a higher \gls{rms} error as they are trying to maximally compensate the \gls{ttl} contribution.

The high accuracy of parameter estimation makes it possible to fully subtract the maneuver signal from the total interferometric output (c.f. Fig.~\myhyperref{fig:ttl-full-man-subtract}). This is especially important in case these maneuvers are going to be used during science operation of the \gls{lisa} mission.

In general, the \gls{ls} estimator has been shown to perform sufficiently well. However, in case of bad signal-to-noise ratios in the \gls{dws} output, significant bias contributions have been found. There can also be strong correlations within the estimated parameters, but this can be mitigated with introducing a regularization term.

All these conclusions were drawn without the inclusion of \gls{gw} signals. Previous studies have already investigated their effects on the inference \cite{hartig2025postprocessing,hartig2025tilt}. With the realistic jitters used here, the \gls{ttl} contribution is limited to a frequency band which makes filtering of the data already necessary. In \cite{hartig2025tilt}, most \gls{gw} events had no influence on the inference from the \num{1}-day dataset. Particular massive black hole binary mergers managed to reduce inference accuracy, but these events could either be removed from the time series or by the application of dedicated filters. Alternatively, in \cite{wang2025postprocessing,fang2025tilttolengthnoisesubtractionpointing} it was shown that a \lq null\rq\ \gls{tdi} channel could be used to estimate the \gls{ttl} coupling coefficients, which suppresses the effects of \glspl{gw} on the inference.

For future work, it would be interesting to include long-term variations for the considered jitters, and upgrade the control loop to the latest understanding of the actual \gls{lisa} satellites. This would then also include modeling \gls{sc} actuation cross-talk between forces and torques, and a more realistic model for \gls{mosa} actuation. Additionally, other maneuver designs, such as \cite{houba_maneuver_gw_22}, could be implemented in this closed-loop settings, to check whether predictions hold up. This can then be used to evaluate the maneuvers not only on the basis of their effectiveness for estimating \gls{ttl} coupling coefficients, but also in terms of saving fuel and impact on science operations. For such longer time simulations it would also be important to model a low-frequency roll-up in the sensing and actuation noises (c.f. Table~\myhyperref{table:noises}).

Furthermore, focusing on the inference pipeline, it is interesting to see which approach (e.g. \gls{ls} vs. MCMC) works better, and how the results depend on the chosen \gls{tdi} channels. This work showed an invariance of the \gls{ls} estimator under orthogonal transformations of the Michelson channels, i.e., using $(A,E,T)$ instead of $(X,Y,Z)$ makes no difference. However, using for example only the \lq null\rq\ channels like in \cite{wang2025postprocessing,muratore_null_23,fang2025tilttolengthnoisesubtractionpointing} would be interesting to explore in this setting as well.

\begin{acknowledgments}
This work was funded in part by the Deutsche Forschungsgemeinschaft (DFG, German Research Foundation) under Germany's Excellence Strategy EXC 2181/1 - 390900948 (the Heidelberg STRUCTURES Excellence Cluster).\\
The authors acknowledge support by the state of Baden-Württemberg through bwHPC. \\
SP gratefully acknowledges support by Deutsches Zentrum für Luft- und Raumfahrt (DLR) with funding of the Bundesministerium für Wirtschaft und Energie with a decision of the Deutsche Bundestag (DLR project reference number FKZ 50OQ2301, based on funding from FKZ 50 OQ 1801). HI thanks the Belgian Federal Science Policy Office (BELSPO) for the provision of financial support in the framework of the PRODEX Programme of the European Space Agency (ESA) under contract number PEA4000144253. \\
The authors thank the LISA Simulation Working Group and the LISA Simulation Expert Group for the lively discussions on all simulation-related activities. \\
Thanks to Marie-Sophie Hartig and Jean-Baptiste Bayle for helpful discussions and support.
\end{acknowledgments}

\section*{Data Availability}
The data that support the findings of this article are not publicly available because of legal restrictions preventing unrestricted public distribution. The data are available from the authors upon reasonable request.

\appendix
\section{Implementation Details} \label{app:lisanode}
\begin{table*}[!ht]
\setlength{\tabcolsep}{12pt}
\renewcommand{\arraystretch}{1.1}
\centering
 \begin{tabular}{l c c c c} 
 \hline\hline
 No. & Sensing channel & Noise floor & Actuation channel & Noise floor \\ 
 \hline 
 1 \vphantom{\Large l} & $x_1^\text{ifo}$/$x_2^\text{ifo}$ & \SI[power-half-as-sqrt,per-mode=symbol]{1.42e-12}{\m\per\Hz\tothe{0.5}} & Thrust $X$ & \SI[power-half-as-sqrt,per-mode=symbol]{2.2e-7}{\newton\per\Hz\tothe{0.5}} \\ 
 2 & $\eta_1^\text{ifo}$/$\eta_2^\text{ifo}$ & \SI[power-half-as-sqrt,per-mode=symbol]{2.0e-9}{\radian\per\Hz\tothe{0.5}} & Thrust $Y$ & \SI[power-half-as-sqrt,per-mode=symbol]{1.3e-7}{\newton\per\Hz\tothe{0.5}} \\
 3 & $\phi_1^\text{ifo}$/$\phi_2^\text{ifo}$ & \SI[power-half-as-sqrt,per-mode=symbol]{2.0e-9}{\radian\per\Hz\tothe{0.5}} & Thrust $Z$ & \SI[power-half-as-sqrt,per-mode=symbol]{3.6e-7}{\newton\per\Hz\tothe{0.5}} \\
 4 & $x_1^\text{grs}$/$x_2^\text{grs}$ & \SI[power-half-as-sqrt,per-mode=symbol]{1.8e-9}{\m\per\Hz\tothe{0.5}} & Thrust $\Theta$ & \SI[power-half-as-sqrt,per-mode=symbol]{7.7e-8}{\newton\m\per\Hz\tothe{0.5}} \\
 5 & $y_1^\text{grs}$/$y_2^\text{grs}$ & \SI[power-half-as-sqrt,per-mode=symbol]{1.8e-9}{\m\per\Hz\tothe{0.5}} & Thrust $H$ & \SI[power-half-as-sqrt,per-mode=symbol]{6.9e-8}{\newton\m\per\Hz\tothe{0.5}} \\
 6 & $z_1^\text{grs}$/$z_2^\text{grs}$ & \SI[power-half-as-sqrt,per-mode=symbol]{3.0e-9}{\m\per\Hz\tothe{0.5}} & Thrust $\Phi$ & \SI[power-half-as-sqrt,per-mode=symbol]{1.3e-7}{\newton\m\per\Hz\tothe{0.5}} \\
 7 & $\theta_1^\text{grs}$/$\theta_2^\text{grs}$ & \SI[power-half-as-sqrt,per-mode=symbol]{120e-9}{\radian\per\Hz\tothe{0.5}} & $f_y^\text{grs}$ & \SI[power-half-as-sqrt,per-mode=symbol]{6.0e-15}{\newton\per\Hz\tothe{0.5}} \\
 8 & $\eta_1^\text{dws}$/$\eta_2^\text{dws}$ & \SI[power-half-as-sqrt,per-mode=symbol]{0.2e-9}{\radian\per\Hz\tothe{0.5}} & $f_z^\text{grs}$ & \SI[power-half-as-sqrt,per-mode=symbol]{1.0e-14}{\newton\per\Hz\tothe{0.5}} \\
 9 & $\phi_1^\text{dws}$/$\phi_2^\text{dws}$ & \SI[power-half-as-sqrt,per-mode=symbol]{0.2e-9}{\radian\per\Hz\tothe{0.5}} & $t_\theta^\text{grs}$ & \SI[power-half-as-sqrt,per-mode=symbol]{1.0e-15}{\newton\m\per\Hz\tothe{0.5}} \\
 10 & & & $t_\eta^\text{grs}$ & \SI[power-half-as-sqrt,per-mode=symbol]{1.0e-15}{\newton\m\per\Hz\tothe{0.5}} \\
 11 & & & $t_\phi^\text{grs}$ & \SI[power-half-as-sqrt,per-mode=symbol]{1.0e-15}{\newton\m\per\Hz\tothe{0.5}} \\
 12 & & & $t_\phi^\text{mosa}$ & \SI[power-half-as-sqrt,per-mode=symbol]{4.5e-14}{\newton\m\per\Hz\tothe{0.5}} \\
 \hline\hline
 \end{tabular}
 \label{table:noises}
 \caption{Table of noises implemented in the closed-loop dynamics, i.e., for sensing and actuation. All noises here are implemented as white noises for now, as a low frequency roll-up would not impact the bandpass-filtered data used for \gls{ttl} parameter inference. Settings largely follow the previous publication \cite{inchauspe23_dynamics}. The acronyms used are local optical interferometry systems (IFOs) for the \gls{tm} displacement sensing along the telescope axis, and Gravitational Reference System (GRS) for electrostatic sensing and actuation of the \glspl{tm}.}
\end{table*}
The simulation was performed using the LISANode suite \cite{bayle2023unified,bayle_lisanode}. The inputs are an orbit file, containing the precomputed orbits, and frequency-planning files \cite{fplan_lisa_24}. The orbits files are produced using LISAOrbits \cite{martens2021trajectory,bayle_lisaorbit}, using the \gls{esa} trailing orbits. The start time for all simulations is then $t_0=0\,\unit{\s}$, which corresponds to $\text{UNIX}\, 2112393530.8167093$, roughly the 9th of December 2036 \cite{bayle_lisaorbit}. The corresponding frequency-planning files were produced from an internal software package, based on a tool called \lq fplan\rq\ \cite{heinzel_fplan_18}, which encode the laser locking scheme. Here, the lock was \lq N1-12\rq, where \lq N1\rq\ refers to the first of the six available locking topologies, and \lq 12\rq\ specifies that the laser on \gls{sc} 1 pointing to 2 is locked to a reference cavity.

Within LISANode, nominal settings were used for noises, such as backlink noise (\SI[power-half-as-sqrt,per-mode=symbol]{3e-12}{\m\per\Hz\tothe{0.5}}, with a $1/f^{2}$ slope below \SI{2}{\milli\Hz}), \gls{tm} acceleration noise (\SI[power-half-as-sqrt,per-mode=symbol]{2.4e-15}{\m\per\s\tothe{2}\per\Hz\tothe{0.5}}, with a $1/f$ slope below \SI{0.4}{\milli\Hz}), interferometer sensing noises (\SI[power-half-as-sqrt,per-mode=symbol]{6.35e-12}{\m\per\Hz\tothe{0.5}}, \SI[power-half-as-sqrt,per-mode=symbol]{1.42e-12}{\m\per\Hz\tothe{0.5}}, \SI[power-half-as-sqrt,per-mode=symbol]{3.32e-12}{\m\per\Hz\tothe{0.5}} for science, \gls{tm}, and reference interferometer respectively, with a $1/f^{2}$ slope below \SI{2}{\milli\Hz}), and laser frequency noise (\SI[power-half-as-sqrt,per-mode=symbol]{30}{\Hz\per\Hz\tothe{0.5}}). Clock, telescope path length, and pseudo-ranging noises were not added. Other sensing and actuation noise from the closed-loop dynamics are shown in Table~\myhyperref{table:noises} and are in part updated from \cite{inchauspe23_dynamics}. The \gls{dws} readout noise levels were adjusted as needed for the different simulation scenarios. Additionally, an actuation noise was introduced for the \gls{mosa} $\phi$ actuation, called $t_\phi^\text{mosa}$ in the table. The level was chosen such that the resulting jitter in \gls{mosa} $\phi$ did not exceed the flat requirement of \SI[power-half-as-sqrt,per-mode=symbol]{2}{\nano\radian\per\Hz\tothe{0.5}} (c.f. Figs.~\myhyperref{fig:jitters-state},\myhyperref{fig:jitters-sensing}).

Note that the \gls{dws} readout noise level specified here is for the phase related variables, and the outputs $\dot{\phi}_{ij}^\text{dws}$, $\dot{\eta}_{ij}^\text{dws}$ will have the derivative of white noises as noise contribution. We will still refer to the amplitude of the white noise \gls{asd} for simplicity when reporting the noise level. Additionally, the value for the \gls{dws} noise \gls{asd} amplitude comes from the noise of the quadrant photodiode (\SI[power-half-as-sqrt,per-mode=symbol]{70}{\nano\radian\per\Hz\tothe{0.5}}) divided by the magnification factor of \num{335}. This would then also imply that consistent $\eta_i^\text{ifo}$, $\phi_i^\text{ifo}$ noise levels should rather be $\SI[power-half-as-sqrt,per-mode=symbol]{70}{\nano\radian\per\Hz\tothe{0.5}}/2.5 \approx \SI[power-half-as-sqrt,per-mode=symbol]{28}{\nano\radian\per\Hz\tothe{0.5}}$. The value given in the table was used to provide consistency with \cite{inchauspe23_dynamics}, but the higher values were checked as well and neither led to significantly different \gls{mosa} or \gls{sc} jitter nor affected the \gls{ttl} coefficient estimation.

Solving the non-linear \gls{eom} for the dynamics was implemented with the Runge-Kutta-Fehlberg 4 algorithm, to ensure sufficient accuracy. This slows down the computational speed, but the \SI{e5}{\s} dataset can still be created in $\sim$\SI{18}{\min}. This assumes the LISANode compilation to C++ was done with O1 optimization, and the application was then run on a AMD EPYC 9454 CPU. Note that the application is single threaded, so no large differences between different computers are expected.

LISANode simulates the internal physics at \SI{16}{\Hz}. All measurements that will also be part of the actual output in the satellites later are fed through an \gls{adc}, which simulates the down-sampling to the final \SI{4}{\Hz} via an anti-aliasing filter. This also introduces a time-delay of the final output signal of about \SI{50}{\s} with respect to the output of the internal states.

The \glspl{asd} in this work have been estimated using an internal tool called 'psd' \cite{bayle_psd}, which uses Welch's method for estimating the \gls{psd} based on the Fast-Fourier Transformation \cite{welch2003use}. The reported parameter $N_\text{avg.}$ specifies how often the algorithm subdivides a given dataset, computes the \gls{psd} on each subset, and then averages over it. The software uses 'nutall4' as the default window \cite{heinzel2002spectrum,nuttall2003some}, which was also used throughout this work.

Generally, this work has greatly benefited from the available software packages NumPy \cite{harris2020array}, SciPy \cite{2020SciPy-NMeth}, and MatplotLib \cite{Hunter:2007}, either directly or indirectly through other packages. Additionally, the Mathematica suite has been helpful for symbolic computations \cite{mathematica}.

\section{Filtering} \label{app:filtering}
\begin{figure*}[!ht]
\centerline{\includegraphics[width=\linewidth, clip]{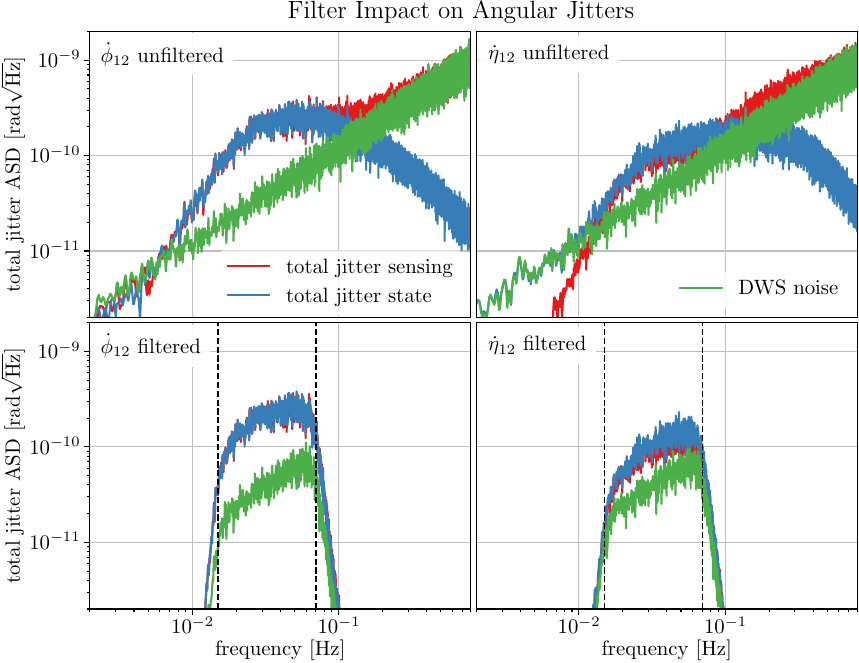}}
\caption{The panels show the impact of the bandpass filters and how the frequency band was chosen. The top panels show the $\phi$ and $\eta$ total \gls{mosa} jitters, i.e., the \gls{dws} measurement outputs for \gls{mosa} 12, split up into the internal state and the readout noise. The bottom panels show the jitters after applying the bandpass filter with frequencies \SI{15}{\milli\Hz} and \SI{70}{\milli\Hz}. These cut-offs were chosen such that a broad frequency range is selected with a good signal-to-noise ratio. The \glspl{asd} are calculated from a \SI{e5}{\s} dataset with $N_\text{avg.}=5$.}
\label{fig:filtering}
\end{figure*}
The parameter inference is based on time-domain data. Looking at Fig.~\myhyperref{fig:jitters-sensing} it is clear that the time-domain signal is completely dominated by the high-frequency \gls{dws} readout noise. Thus these outputs must be filtered before any inference in the time domain is possible.

Here a Butterworth filter of order $N=5$ is chosen, as this has the flattest possible spectral shape in the bandpass region \cite{Bianchi2007ElectronicFS}. As we have complete information within the simulation, a lower and upper frequency cut-off can be chosen to maximize the available signal-to-noise ratio in the \gls{dws} output. The final values are \SI{15}{\milli\Hz} and \SI{70}{\milli\Hz} and can be viewed in Fig.~\myhyperref{fig:filtering}. These values are not optimized beyond this, as in the actual mission a precise split between jitter and noise is not possible. However, a rough frequency range for the bandpass filter can be taken directly from the \gls{dws} outputs.

The filter choice introduces significant ringing at the beginning and end of the time series. This is cut away with a margin of \SI{1000}{\s} for datasets without maneuvers, and \SI{750}{\s} for those with, from the time series data. These margins do not count towards the reported dataset length.

\section{Bias for Least-Square Estimator} \label{app:bias}
\begin{figure*}[!ht]
\centerline{\includegraphics[width=\linewidth, clip]{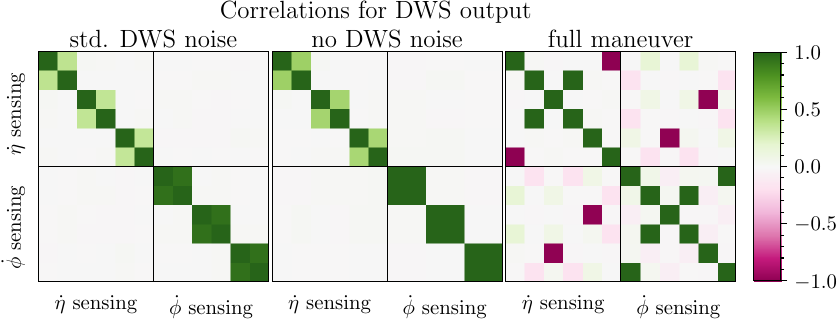}}
\caption{The panels show the normalized covariances, i.e., correlations, between the \gls{dws} channels in a single simulation. The channels are order as $\{12,13,23,21,31,32\}$. The first panel is the standard simulation with \gls{dws} readout noise at \SI[power-half-as-sqrt,per-mode=symbol]{0.2}{\nano\radian\per\Hz\tothe{0.5}}. The second is a simulation without any \gls{dws} noise. The last panel looks at the situation when full maneuvers are present. The first two panels show how removing the \gls{dws} noise increases the correlation between the $\phi$ \gls{dws} channels of the same \gls{sc}. The corresponding $\eta$ correlations are weaker and remain largely unaffected by the \gls{dws} noise level. When introducing maneuvers, new strong (anti-)correlations are present, due to the design of the maneuver. Figure~\myhyperref{fig:ls-correlations} shows that these correlations are not carried over into the \gls{ttl} coefficient estimates.}
\label{fig:ls-dws-correlations}
\end{figure*}
\begin{figure*}[!ht]
\centerline{\includegraphics[width=\linewidth, clip]{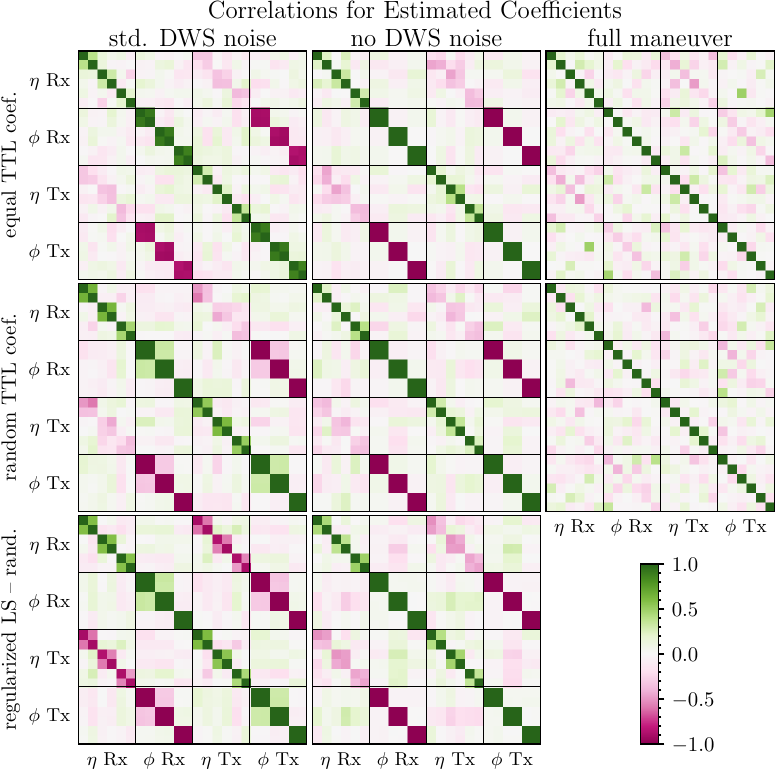}}
\caption{The panels show the normalized covariances, i.e., correlations, between the estimated \gls{ttl} coefficient deviations of \num{100} simulations, i.e., correlations of $\hat{C}_\text{TTL}-C_\text{TTL}$. For each parameter subset, the order is $\{12,13,23,21,31,32\}$. The top row looks at the case of equal \gls{ttl} coefficients in the simulation, the middle for coefficients drawn from a uniform distribution. The bottom row investigates the output of the regularized \gls{ls} estimator for random \gls{ttl} coefficients. The first column is the standard simulation with \gls{dws} readout noise at \SI[power-half-as-sqrt,per-mode=symbol]{0.2}{\nano\radian\per\Hz\tothe{0.5}}. The second column is a simulation without any \gls{dws} noise. The last column looks at the situation when full maneuvers are present.\\ 
The panels show that the maneuvers manage to get rid of cross-correlations between coefficients, while the regular simulations have quite strongly coupled $\phi$ coefficients. The regularization scheme does not alleviate these, it can actually introduce additional correlations for $\eta$.}
\label{fig:ls-correlations}
\end{figure*}
This section calculates the bias of the \gls{ls} estimator, extending previous work \cite{hartig2025tilt} to the case of non-white \gls{dws} readout noise, and differing \gls{sc} $\Theta$ and $H$ jitter. 

The bias is defined as 
\begin{equation}
    \operatorname{bias}(\hat{\myvec{C}}_\text{TTL}) = \lim_{N\rightarrow\infty} \hat{\myvec{C}}_\text{TTL} - \myvec{C}_\text{TTL}\ ,
\end{equation}
with $N$ being the number of elements in the time series in the definition of $\hat{\myvec{C}}_\text{TTL}$ (c.f. Eq.~\myhyperref{eq:ls-solution}). As a starting point, use the lowest order approximations for the \gls{dws} outputs from Eqs.~\myhyperref{eq:dws-approx-eta1}-\myhyperref{eq:dws-approx-eta2-equal}. Following \cite{hartig2025tilt}, we are looking at the unequal arm length configuration for the correlations, but approximately work with a breathing angle $\varphi_b=\pi/3$. 

The relevant variables are the \gls{sc} $i\in\{1,2,3\}$ angular velocities $(\omega_i^\Theta,\omega^H_i,\omega^\Phi_i)$, the \gls{mosa} angular velocities $(\omega_{ij}^\eta,\omega_{ij}^\phi)$, and \gls{dws} readout noise terms $(\dot{n}_{ij}^\eta,\dot{n}_{ij}^\phi)$ for $j\in\{1,2,3\}$, $i\neq j$. We assume that the spectral shapes are identical between different \gls{sc} and \glspl{mosa}, but not necessarily between different types (e.g. $\omega^\Theta$ and $\omega^\Phi$).

The terms that come up in the calculation then have the form 
\begin{equation}
    \lim_{N\rightarrow\infty} \frac{1}{N-1} \left(\omega^\Phi_i\right)^T \omega^\Phi_j = \delta_{ij}\sigma^{2}_\Phi\ ,
\end{equation}
with the Kronecker symbol $\delta_{ij}$ giving zero for separate \gls{sc}, and $\sigma^{2}_\Phi$ the variance of the \gls{sc} $\Phi$ jitter (for non-zero mean, this needs to be additionally subtracted). 

As the jitters considered are not white, the correlation of the jitter with a time-delayed version of itself is non-zero. For this, introduce the normalized auto-correlations via
\begin{equation}
    \rho_\alpha^{k\tau} \coloneqq \lim_{N\rightarrow\infty} \frac{1}{(N-1)\sigma^{2}_\alpha} \left(\omega^\alpha\right)^T \dot{D}_{k\tau} \omega^\alpha \ ,
\end{equation}
with $k\in\{2,4,6,8\}$, $\alpha\in\{\Theta,H,\Phi,\eta,\phi,n_\eta,n_\phi\}$ and $\tau$ the average light travel time along one arm. Working with unequal arm lengths in \gls{lisa} then means that e.g. 
\begin{equation}
\lim_{N\rightarrow\infty} \frac{1}{N-1} \left(\dot{D}_{121}\omega^\Phi_1\right)^T \left(\dot{D}_{131}\omega^\Phi_1\right) = 0\ .
\end{equation}

Then the bias formula can be written out in terms of the basic variables and time delays, using e.g.~the appendix of \cite{wegener2025design}. The computations can be performed with tools like \cite{mathematica}, but it is helpful to note that correlations of \gls{sc} jitter, \gls{mosa} jitter, and readout noise between different \gls{sc} are zero. So always the four dimensional subspace, e.g.~$(C_{12\eta}^\text{Rx},C_{13\eta}^\text{Rx},C_{12\eta}^\text{Tx},C_{13\eta}^\text{Tx})$ can be considered (this works also for the matrix inversion, which receives a block-diagonal structure).

The results is then similar to the original \cite{hartig2025tilt}, but extended by the auto-correlations of the \gls{dws} readout noise. Using indices $\{i,j,k\}=\{1,2,3\}$, the bias is given by
\begin{align}
    \begin{split}
    \operatorname{bias}(\hat{C}_{ij\alpha}^\text{Rx}) = &\,K_{\alpha 1} (C_{ij\alpha}^\text{Rx} + C_{ik\alpha}^\text{Rx} - C_{ij\alpha}^\text{Tx} - C_{ik\alpha}^\text{Tx}) \\
    &+K_{\alpha 2} (C_{ij\alpha}^\text{Rx} - C_{ik\alpha}^\text{Rx} - C_{ij\alpha}^\text{Tx} + C_{ik\alpha}^\text{Tx}) \\
    &+K_{\alpha 3} (C_{ij\alpha}^\text{Rx} + C_{ik\alpha}^\text{Rx} + C_{ij\alpha}^\text{Tx} + C_{ik\alpha}^\text{Tx}) \\
    &+K_{\alpha 4} (C_{ij\alpha}^\text{Rx} - C_{ik\alpha}^\text{Rx} + C_{ij\alpha}^\text{Tx} - C_{ik\alpha}^\text{Tx})\, ,
    \end{split} \label{eq:bias-general-rx}\\
    \begin{split}
    \operatorname{bias}(\hat{C}_{ij\alpha}^\text{Tx}) = &-K_{\alpha 1} (C_{ij\alpha}^\text{Rx} + C_{ik\alpha}^\text{Rx} - C_{ij\alpha}^\text{Tx} - C_{ik\alpha}^\text{Tx}) \\
    &-K_{\alpha 2} (C_{ij\alpha}^\text{Rx} - C_{ik\alpha}^\text{Rx} - C_{ij\alpha}^\text{Tx} + C_{ik\alpha}^\text{Tx}) \\
    &+K_{\alpha 3} (C_{ij\alpha}^\text{Rx} + C_{ik\alpha}^\text{Rx} + C_{ij\alpha}^\text{Tx} + C_{ik\alpha}^\text{Tx}) \\
    &+K_{\alpha 4} (C_{ij\alpha}^\text{Rx} - C_{ik\alpha}^\text{Rx} + C_{ij\alpha}^\text{Tx} - C_{ik\alpha}^\text{Tx})\, ,
    \end{split} \label{eq:bias-general-tx}
\end{align}
with $\alpha\in\{\eta,\phi\}$. The coefficients are given by
\begin{align}
    K_{\eta 1} &= \frac{\operatorname{AC}_{n_\eta,1}}{4\operatorname{AC}_{\eta,1} + 4\operatorname{AC}_{n_\eta,1} +  2\operatorname{AC}_{\Theta,1}} \\
    K_{\eta 2} &= \frac{\operatorname{AC}_{n_\eta,1}}{4\operatorname{AC}_{\eta,1} + 4\operatorname{AC}_{n_\eta,1} + 6\operatorname{AC}_{H,1}} \\
    K_{\eta 3} &= \frac{\operatorname{AC}_{n_\eta,2}}{4\operatorname{AC}_{\eta,2} + 4\operatorname{AC}_{n_\eta,2} +  2\operatorname{AC}_{\Theta,4} +  12\operatorname{AC}_{H,3}} \\
    K_{\eta 4} &= \frac{\operatorname{AC}_{n_\eta,2}}{4\operatorname{AC}_{\eta,2} + 4\operatorname{AC}_{n_\eta,2}  +  4\operatorname{AC}_{\Theta,3} +  6\operatorname{AC}_{H,4}}\\
    K_{\phi 1} &= \frac{\operatorname{AC}_{n_\phi,1}}{4\operatorname{AC}_{\phi,1} + 4\operatorname{AC}_{n_\phi,1} } \\
    K_{\phi 2} &= \frac{\operatorname{AC}_{n_\phi,1}}{4\operatorname{AC}_{\phi,1} + 4\operatorname{AC}_{n_\phi,1} + 8\operatorname{AC}_{\Phi,1}} \\
    K_{\phi 3} &= \frac{\operatorname{AC}_{n_\phi,2}}{4\operatorname{AC}_{\phi,2} + 4\operatorname{AC}_{n_\phi,2} + 16\operatorname{AC}_{\Phi,3}} \\
    K_{\phi 4} &= \frac{\operatorname{AC}_{n_\phi,2}}{4\operatorname{AC}_{\phi,2} + 4\operatorname{AC}_{n_\phi,2} + 8\operatorname{AC}_{\Phi,4}}
\end{align}
using
\begin{equation}
    \operatorname{AC}_{\alpha,1} = \sigma_{\alpha}^{2}(3-4\rho_{\alpha}^{2\tau}- 4\rho_{\alpha}^{4\tau}+ 4\rho_{\alpha}^{6\tau}-\rho_{\alpha}^{8\tau})\ ,
\end{equation}
for $\alpha\in\{\Theta,H,\Phi,\eta,\phi,n_\eta,n_\phi\}$ and
\begin{equation}
    \operatorname{AC}_{\alpha,2} = \sigma_{\alpha}^{2}(13-4\rho_{\alpha}^{2\tau}- 12\rho_{\alpha}^{4\tau}+ 4\rho_{\alpha}^{6\tau}+\rho_{\alpha}^{8\tau})\ ,
\end{equation}
for $\alpha\in\{\eta,\phi,n_\eta,n_\phi\}$. Furthermore, define
\begin{align}
    \operatorname{AC}_{\alpha,3} &= \sigma_{\alpha}^{2}(3-\rho_{\alpha}^{2\tau}- 2\rho_{\alpha}^{4\tau}+ \rho_{\alpha}^{6\tau})\ , \\
    \operatorname{AC}_{\alpha,4} &= \sigma_{\alpha}^{2}(7-2\rho_{\alpha}^{2\tau}- 8\rho_{\alpha}^{4\tau}+ 2\rho_{\alpha}^{6\tau}+\rho_{\alpha}^{8\tau})\ 
\end{align}
for $\alpha\in\{\Theta,H,\Phi\}$.

Dividing out the noise terms in the coefficient formulas then shows that the bias can be expressed in terms of the signal-to-noise ratio in the \gls{dws} outputs. This interpretation is not exactly true in the case of non-white jitter, as the auto-correlations complicate this case.

The extended simulation output can be used to calculate these prefactors $K_\eta$, $K_\phi$ approximately. Here are the results for nominal \gls{dws} noise settings of \SI[power-half-as-sqrt,per-mode=symbol]{0.2}{\nano\radian\per\Hz\tothe{0.5}}. Note that for equal \gls{ttl} coefficients, only $K_{\eta 3}$, $K_{\phi 3}$ contribute to the bias. To calculate the values, the filtered versions of the jitters and noises are used (c.f. App.~\myhyperref{app:filtering}). The given coefficients here are unitless as ratios of auto-correlations
\begin{align}
    &K_{\eta 1} \approx -0.0431\ , &&K_{\phi 1} \approx -0.2500\ , \label{eq:bias-coeff1-explicit}\\
    &K_{\eta 2} \approx -0.0229\ , &&K_{\phi 2} \approx -0.0037\ , \label{eq:bias-coeff2-explicit}\\
    &K_{\eta 3} \approx 0.0378\ , &&K_{\phi 3} \approx 0.0122\ , \label{eq:bias-coeff3-explicit}\\
    &K_{\eta 4} \approx 0.0369\ , &&K_{\phi 4} \approx 0.0108\ .\label{eq:bias-coeff4-explicit}
\end{align}

\section{Estimated Uncertainties}\label{app:error}
It is also possible to estimate the errors of individual parameters with the following scheme \cite{wegener2025design}: in the idealized case of no \gls{dws} readout noise (i.e., no bias), and assuming that instrumental noise results from a Gaussian process, the covariance matrix of the estimated parameters is given by
\begin{equation}
    \hat{\mymat{C}} = \sigma^2_\text{inst.}\left({\dot{\myvec{\alpha}}_\text{TDI}}^T \cdot \dot{\myvec{\alpha}}_\text{TDI}\right)^{-1}\ ,
\end{equation}
with $\sigma^2_\text{inst.}$ the variance of the instrumental noise, i.e., the variance of the full interferometric output without any \gls{ttl} contribution and no \gls{dws} readout noise. Estimating this from a \SI{e5}{\second} dataset using a bandpass filter (c.f. App.~\myhyperref{app:filtering}) gives \SI{1.73}{\micro\hertz}. The standard deviations of the estimated parameters is then given by
\begin{equation}
    \myvec{\sigma}_\text{TTL} = A_\text{heur.} \sigma_\text{inst.}\operatorname{diag}\!\left(\sqrt{\left({\dot{\myvec{\alpha}}_\text{TDI}}^T \cdot \dot{\myvec{\alpha}}_\text{TDI}\right)^{-1}}\,\right) \label{eq:ttl-coeff-err}
\end{equation}
where $\sqrt{\dots}$ is understood as an element-wise operation and $A_\text{heur.}$ is a heuristic factor taking into account that usual inference does not fulfill the idealized conditions of this calculation. However, when performing similar efforts as \cite{wegener2025design}, we find $A_\text{heur.}\approx 10$. The setting here seems to be too far away from the idealizing assumptions made due to working with frequencies vs. phases, having a large bias contribution, and  using bandpass filters.

Thus, for the error bars displayed in Figs.~\myhyperref{fig:ttl-inference-result},\myhyperref{fig:ttl-inference-result-no-dws},\myhyperref{fig:ttl-inference-result-man} are calculated from the standard deviation of \num{100} simulations in the respective scenarios. 

\section{Correlations}\label{app:corr}
This sections provides details about the correlations in the inferred \gls{ttl} coefficients. A comprehensive plot of the different scenarios can be found in Fig.~\myhyperref{fig:ls-correlations}. 

However, starting with Fig.~\myhyperref{fig:ls-dws-correlations}, it is evident that for the \num{1}-day datasets without maneuvers, there are already a lot of correlations present between the different \gls{dws} outputs, especially for the $\phi$ channels. Moving then to the case of no \gls{dws} readout noise, the $\phi$ correlations become stronger. The maneuver dataset also exhibits strong correlations, but they were specifically chosen such that after \gls{tdi} they are not present anymore \cite{wegener2025design}.

This picture is carried over when looking at the full correlations in the inferred parameters (c.f. Fig.~\myhyperref{fig:ls-correlations}): these are correlations
\begin{equation}
    \operatorname{corr}(\myvec{v})_{ij} \coloneqq \frac{\mymat{C}_{ij}}{\sqrt{\mymat{C}_{ii}}\sqrt{\mymat{C}_{jj}}} \in [-1,1]\subset \mathbb{R}
\end{equation}
of the vector $\myvec{v}=\hat{\myvec{C}}_\text{TTL}-\myvec{C}_\text{TTL}$ with covariance matrix $\mymat{C}=\operatorname{cov}(\myvec{v})$. The figure shows the resulting for the outputs of \num{100} simulations for each scenario considered.

The upper-left plot shows the correlations for a \num{1}-day dataset with nominal \gls{dws} readout noise and equal \gls{ttl} coefficients. These correlations are due to the correlations present in the \gls{dws} outputs shown before. This pattern was also reported in \cite{wegener2025design,hartig2025tilt,paczkowski_postprocessing_2022}.
\begin{figure*}[!ht]
\centerline{\includegraphics[width=\linewidth, clip]{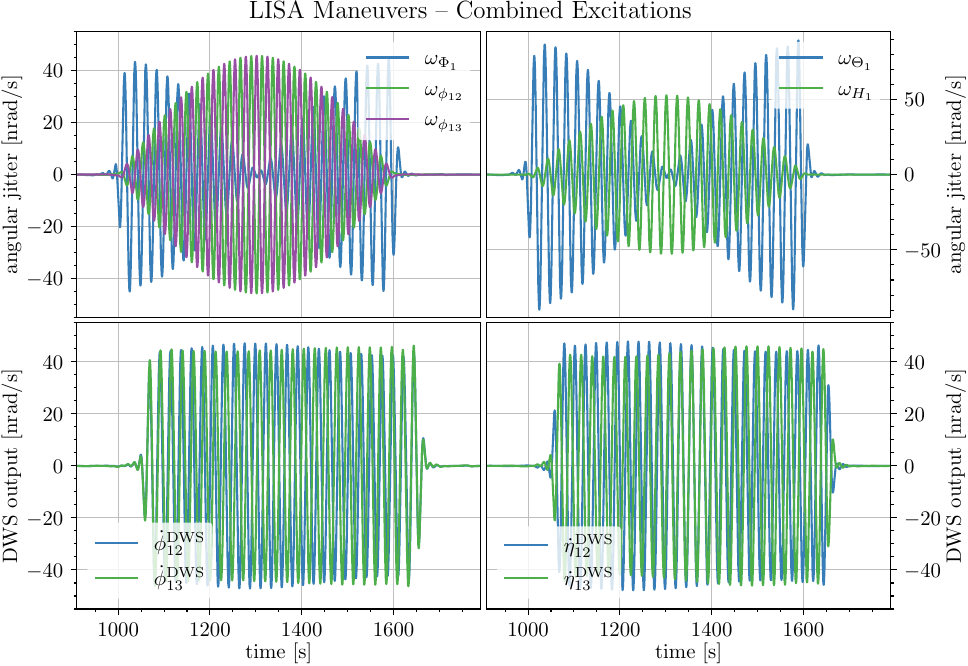}}
\caption{The panels show how $\phi$ (left) and $\eta$ (right) maneuvers are induced in both \glspl{mosa} on a single \gls{sc} at different frequencies \SI[parse-numbers=false]{43.\bar{3}}{\milli\Hz} and \SI[parse-numbers=false]{41.\bar{6}}{\milli\Hz}. This can be used to construct a full type 2 maneuver, with exciting all $\phi$ channels in phase \num{1}, and then all the $\eta$ channels in phase \num{2}. The top panels display the actual rotations induced in the \gls{sc}/\gls{mosa} degrees-of-freedom. The bottom panels show the resulting \gls{dws} readout, with excitations present in both \glspl{mosa}. Details on the required excitations can be found in Sec.~\myhyperref{sec:exp-2-man}. Maneuver duration is \SI{600}{\s}. The shift between top and bottom panels in the start of the excitations is due to the delay caused by the \gls{adc}. A bandpass filter has been applied to the data with frequencies \SI{15}{\milli\Hz} and \SI{70}{\milli\Hz}.}
\label{fig:ttl-full-man2}
\end{figure*}

The upper-middle plot shows the results for the case of no \gls{dws} readout noise. This makes the strong (anti-)
correlations in the $\phi$ channels perfect, i.e., $\pm 1$. This case will be dealt with in the next subsection.

The right column handles the case for the maneuver datasets. There, by construction \cite{wegener2025design}, no correlations are present between inferred parameters. This shows that the theoretical construction of the uncorrelated maneuvers also holds true in the more realistic closed-loop dynamics setting.

Moving to the second row of the figure, now random \gls{ttl} coefficients are drawn from a uniform distribution $\mathcal{U}([-23,23]\,\unit[per-mode=symbol]{\milli\metre\per\radian})$, and $\mathcal{U}([-2.3,2.3]\,\unit[per-mode=symbol]{\milli\metre\per\radian})$ in the maneuver dataset. This seemingly increases the correlations for the \num{1}-day dataset, just like when going to very low levels of \gls{dws} readout noises. This case will also be handled in the next subsections.

The last row shows the correlations for the inferred parameters from the \gls{ls} estimator with Tikhonov regularization, again with randomly drawn \gls{ttl} coefficients. This seems to add correlations between the $\eta$ Rx and $\eta$ Tx channels. Otherwise, this presents the typical picture known from the random \gls{ttl} coefficient case.

\subsection{No DWS readout noise}
Considering the \gls{dws} output without noise, it is given by 
\begin{equation}
    \dot{\phi}^\text{dws}_{ij} = \omega^\Phi_i + \omega^\phi_{ij}
\end{equation}
for \gls{sc} $i \in\{1,2,3\}$, and $j\in\{1,2,3\}$ with $i\neq j$. As filtering of the time series is necessary for the inference, it will also be considered here. In the bandpass frequency range, the \gls{mosa} jitter is subdominant to the \gls{sc} jitter (roughly $4-5$ orders of magnitude). Thus, in this scenario, the \gls{dws} $\phi$ outputs on one \gls{sc} are identical. 

After \gls{tdi}-X, the entries in $\dot{\myvec{\alpha}}_\text{TDI}^\text{dws}$ corresponding to $C^{Rx}_{ij\phi}$, $ C^{Rx}_{ik\phi}$ become the negative of each other, and in \gls{tdi}-Y and \gls{tdi}-Z, $C^{Rx}_{ij\phi}$, $ C^{Rx}_{ik\phi}$ become identical, using the approximation for an equilateral triangle constellation of \gls{lisa} \cite{wegener2025design}. This then means that only specific sums and differences can be estimated together, but the matrix connecting the two descriptions is not invertible \cite{wegener2025design}. The two implications for $\{i,j,k\}=\{1,2,3\}$ are
\begin{enumerate}
    \item $C^{Rx}_{ij\phi} = C^{Rx}_{ik\phi}$
    \item $C^{Rx}_{ij\phi} = -C^{Tx}_{ij\phi}$
\end{enumerate}
which are exactly the correlations that can be seen in Fig.~\myhyperref{fig:ls-correlations}. In terms of the \gls{ls} estimator, this correlation introduces degenerate hyperplanes, i.e., hyperplanes that all have values close to the minimum of the cost function. The \gls{ls} estimator picks a point in the hyperplane, but this can be very far away from the true values. Moving off the hyperplane is still penalized, thus, e.g.~only the sums and differences $C^{Rx}_{12\phi}-C^{Rx}_{13\phi}$, $C^{Tx}_{12\phi}-C^{Tx}_{13\phi}$, $C^{Rx}_{12\phi} +C^{Tx}_{12\phi}$, $C^{Rx}_{13\phi} +C^{Tx}_{13\phi}$ can be estimated well.

\subsection{Random TTL coefficients}
For this case, consider the following: when changing the random \gls{ttl} coefficients between simulations, the underlying simulations are unaffected by the coefficient choices. Thus, when investigating the correlations between $\hat{\myvec{C}}_\text{TTL}-\myvec{C}_\text{TTL}$, this has two components: the underlying correlations stemming from the \gls{dws} output propagated through \gls{tdi}, and additional correlations from the bias of the estimator. 

For the first part, the estimated correlations from the equal \gls{ttl} coefficient case can be used as a proxy: they are unaffected by the constant offset that the bias provides for the random vector of coefficients. 

The added correlations due to the bias can be directly calculated. For this, consider the case $N\rightarrow\infty$, as for the bias. Then, as the underlying simulations are uncorrelated for different realizations, the coefficients $K_\eta$, $K_\phi$ in Eqs.~\myhyperref{eq:bias-general-rx}-\myhyperref{eq:bias-general-tx} can be considered constant. The \gls{ttl} coefficients are independent random variables all drawn from a uniform distribution $\mathcal{U}([a,b])$. Again, like in the bias computation, there are no correlations between the $\eta$ and $\phi$ coefficients, and also not between coefficients of different \gls{sc}. So we can always look at a four-dimensional subsystem. Without restrictions of generality, look at the vectors $\myvec{v}_\alpha=(C^\text{Rx}_{12\alpha},C^\text{Rx}_{13\alpha},C^\text{Tx}_{1\alpha},C^\text{Tx}_{13\alpha})^T$ for $\alpha\in\{\eta,\phi\}$. The covariance matrix $\operatorname{cov}(\myvec{v}_\alpha)$ is then given by
\begin{equation}
    4\sigma^2_\text{TTL}\begin{pmatrix}
        K^{\text{comb.}}_{\alpha 1} & K^{\text{comb.}}_{\alpha 1} & K^{\text{comb.}}_{\alpha 2} & K^{\text{comb.}}_{\alpha 2} \\
        K^{\text{comb.}}_{\alpha 1} & K^{\text{comb.}}_{\alpha 1} & K^{\text{comb.}}_{\alpha 2} & K^{\text{comb.}}_{\alpha 2} \\
        K^{\text{comb.}}_{\alpha 2} & K^{\text{comb.}}_{\alpha 2} & K^{\text{comb.}}_{\alpha 1} & K^{\text{comb.}}_{\alpha 1} \\
        K^{\text{comb.}}_{\alpha 2} & K^{\text{comb.}}_{\alpha 2} & K^{\text{comb.}}_{\alpha 1} & K^{\text{comb.}}_{\alpha 1}
    \end{pmatrix}
\end{equation}
with 
\begin{align}
    K^{\text{comb.}}_{\alpha 1} &= K_{\alpha 1}^2 + K_{\alpha 2}^2 + K_{\alpha 3}^2 + K_{\alpha 4}^2 \ ,\\
    K^{\text{comb.}}_{\alpha 2} &= -K_{\alpha 1}^2 - K_{\alpha 2}^2 + K_{\alpha 3}^2 + K_{\alpha 4}^2\ .
\end{align}

For $\myvec{v}_\phi$ this is dominated by $K_{\phi 1}$, by the relative strength of the coefficients calculated in Eqs.~\myhyperref{eq:bias-coeff1-explicit}-\myhyperref{eq:bias-coeff4-explicit}, so we can make the approximation $K^{\text{comb.}}_{\phi 1}\approx K_{\phi 1}^2$ and $K^{\text{comb.}}_{\phi 2}\approx -K_{\phi 1}^2$, such that the correlation matrix is approximately 
\begin{equation}
    \operatorname{corr}(\myvec{v}_\phi)\approx4\sigma^2_\text{TTL} K_{\alpha 1}^2\begin{pmatrix}
        \phantom{-}1 & \phantom{-}1 & -1 & -1 \\
        \phantom{-}1 & \phantom{-}1 & -1 & -1 \\
        -1 & -1 & \phantom{-}1 & \phantom{-}1 \\
        -1 & -1 & \phantom{-}1 & \phantom{-}1
    \end{pmatrix}\ ,
\end{equation}
with $\sigma^2_\text{TTL}$ the variance of the random coefficients. This shows the familiar pattern from the computed correlations between estimated $\phi$ \gls{ttl} coefficients. This is not a valid approximation for $\myvec{v}_\eta$, which also aligns with the measured correlations for the \gls{ttl} coefficients, which does not see an increase in the $\eta$ correlations going from the equal to the random \gls{ttl} coefficient case. Note also that the strength of the contribution of the bias to the full covariance matrix depends on the interval chosen for the uniform distribution, through the variance of the \gls{ttl} coefficients $\sigma^2_\text{TTL}=\frac{1}{12}(b-a)^{2}$.

\section{Different Maneuver}\label{app:maneuver}
This section investigates several different scenarios from the main results on maneuvers in Sec.~\myhyperref{sec:exp-2-man}. In these maneuvers from \cite{wegener2025design}, the $\phi$ and $\eta$ maneuvers are mixed in the two phases. The possibility of first doing one, then the other is discarded, as this would require higher thruster forces to achieve the same amplitude in the \gls{dws} output. In Fig.~\myhyperref{fig:ttl-full-man2}, the states and sensing outputs can be seen for two different frequency excitations being present at the same time.

\begin{figure}[!ht]
\centerline{\includegraphics[width=\linewidth, clip]{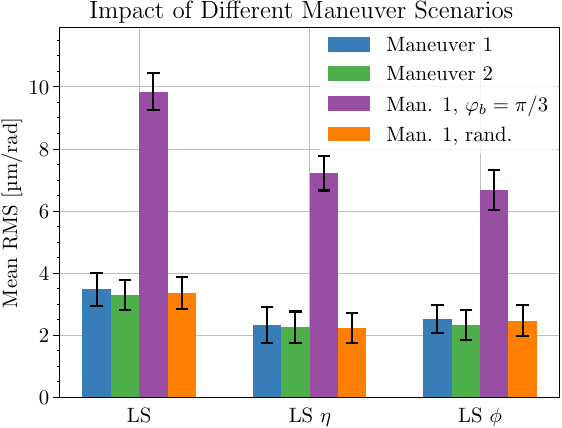}}
\caption{Results of inference when running simulations with maneuvers with different scenarios. Each bar corresponds to the mean of the \gls{ttl} coefficient estimates' \gls{rms} error over \num{100} simulations, with the error bars given by the standard deviation. The result is split up in terms of the contributions of $\eta$ and $\phi$ to the total \gls{rms} error. The four different scenarios are: maneuvers of type \num{1} (mixed $\eta$ and $\phi$ excitations), type \num{2} (separated $\eta$ and $\phi$), maneuver type \num{1} with setting the breathing angle to $\pi/3$ in the \gls{ttl} contribution calculation inside the simulation, and lastly maneuver type \num{1} with randomly sampled \gls{ttl} coefficients from the interval $[-2.3,2.3]\,\unit[per-mode=symbol]{\milli\metre\per\radian}$.
}
\label{fig:ttl-full-man-rand-coeff}
\end{figure}

Using this second type of maneuvers, having first a phase of all $\phi$ maneuvers, then a second phase for $\eta$, this is checked against the regular maneuvers for \num{100} simulations each. As can be seen in Fig.~\myhyperref{fig:ttl-full-man-rand-coeff}, there is no difference within the estimated uncertainties in how close the inferred \gls{ttl} parameters are to the true ones.

The figure contains two additional scenarios for comparison. Firstly, using random \gls{ttl} coefficients drawn from a uniform distribution $\mathcal{U}([-2.3,2.3]\,\text{mm/rad})$ instead of equal \gls{ttl} coefficients of $2.3\,\text{mm/rad}$ makes no real difference for the inference either--the resulting \gls{rms} values are not significantly different. This is expected, as there is no significant bias of the estimator that could add additional correlations into the inferred parameters. 
\begin{figure*}[!ht]
\centerline{\includegraphics[width=\linewidth, clip]{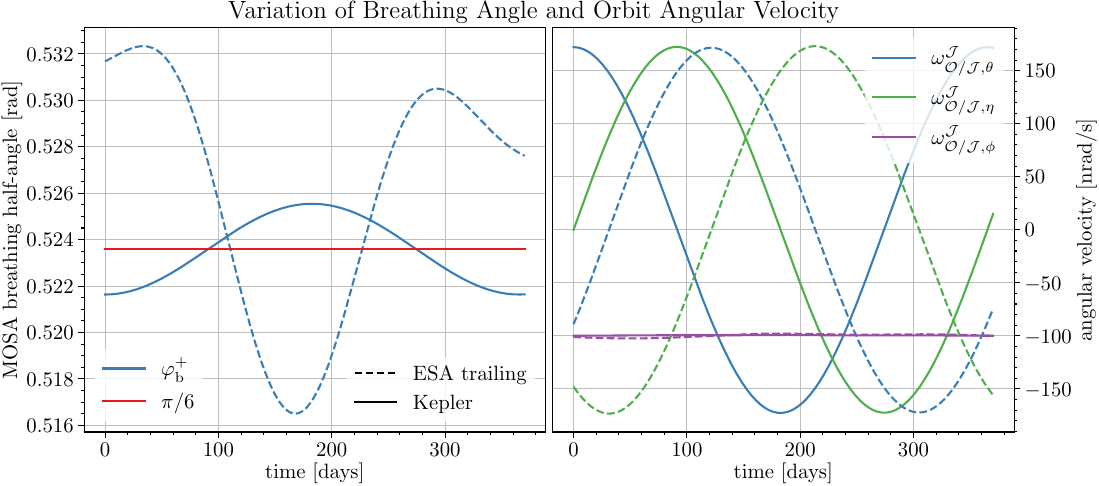}}
\caption{The panels show the breathing angle of the constellation, and the angular velocity induced by the orbit, varying over the course of a year. On the left the positive breathing half-angle $\varphi_b^+$ is shown for two pre-computed orbits: Kepler-type and \gls{esa}-optimized with the \gls{lisa} constellation trailing Earth. The right panel shows the variation of the components of the angular velocity vector induced by the orbital motion.}
\label{fig:long-term-changes}
\end{figure*}

Secondly, a scenario was created in which in the internal modeling of the \gls{ttl} contribution to the interferometer output, the breathing angle was set to $\pi/3$, not the angle computed from the orbits. The \gls{dws} output was still modeled correctly. The impact on the inference is usually believed to be small \cite{hartig2025tilt}, as the deviation from $\pi/3$ is small (c.f. App.~\myhyperref{app:long-term-var}). In the figure it can however be seen that the \gls{rms} value is about three times worse, both in the $\eta$ and $\phi$ parameters, even though the breathing angle only enters the $\eta$ \gls{ttl} contribution to the interferometer. The overall estimation error is still small at $\sim0.4\%$.

\section{Long-Term Variation}\label{app:long-term-var}
Here we provide some insights into the long-term variations captured in the present dynamics modeling. Parameters that currently enter the \gls{eom} are the breathing angles $\varphi_b$ and the orbital velocities/accelerations of the orbit frame with respect to the solar frame $\mathcal{J}$. For the precomputed \lq \gls{esa} trailing\rq\ and \lq Kepler\rq\ orbits \cite{martens2021trajectory,bayle_lisaorbit}, Fig.~\myhyperref{fig:long-term-changes} shows the changes over one year. Note that both these contributions can be directly computed from the orbit files.

As these terms play a smaller role in the \gls{eom}, it is not expected that these effects will cause a drastic effect on the underlying jitters, and thus the scale of the expected \gls{ttl} contribution to the interferometer output. However, for cases in which the \gls{ttl} coefficients need to be inferred accurately, it is important to include the drifting of these parameters.

Effects that are not included in the present dynamics modeling are the changes in the moment-of-inertia of the \gls{sc} due to the movement of the \glspl{mosa} or due to depletion of stored gas because of thruster usage, or effects from deteriorating performance of instrument components.

\bibliography{references}

\end{document}